\documentclass[aps,prd,reprint,superscriptaddress,nofootinbib,longbibliography,floatfix]{revtex4-2}

\usepackage[T1]{fontenc}
\usepackage{graphicx}
\usepackage{epsf}
\usepackage{bm}
\usepackage{epstopdf}
\usepackage{comment}
\usepackage{color}
\usepackage[dvipsnames]{xcolor}
\usepackage{verbatim}
\usepackage{multirow}
\usepackage{soul}
\usepackage{physics}
\usepackage{microtype}
\usepackage{amsmath}
\usepackage{amssymb}
\usepackage[normalem]{ulem}
\usepackage{enumitem}
\usepackage[colorlinks=true,linkcolor=blue,citecolor=blue,urlcolor=blue]{hyperref}
\usepackage[capitalize]{cleveref}
\usepackage{orcidlink}

\makeatletter\let\expandableinput\@@input\makeatother

\def\Mpl{M_{\rm P}}

\DeclareRobustCommand{\VAN}[3]{#2}
\let\VANthebibliography\thebibliography
\def\thebibliography{\DeclareRobustCommand{\VAN}[3]{##3}\VANthebibliography}
\begin{document}

\title{$\Lambda_{\rm s}$CDM cosmology from a type-II minimally modified gravity}

\author{\"{O}zg\"{u}r Akarsu\,\orcidlink{0000-0001-6917-6176}}
\email{akarsuo@itu.edu.tr}
\affiliation{Department of Physics, Istanbul Technical University, Maslak 34469 Istanbul, T\"{u}rkiye}

\author{Antonio De Felice\,\orcidlink{0000-0002-5556-4693}}
\email{antonio.defelice@yukawa.kyoto-u.ac.jp}
\affiliation{Center for Gravitational Physics and Quantum Information, Yukawa Institute for Theoretical Physics, Kyoto University, 606-8502, Kyoto, Japan}

\author{Eleonora Di Valentino\,\orcidlink{0000-0001-8408-6961}}
\email{e.divalentino@sheffield.ac.uk}
\affiliation{School of Mathematics and Statistics, University of Sheffield, Hounsfield Road, Sheffield S3 7RH, United Kingdom}

\author{Suresh Kumar\,\orcidlink{0000-0002-1638-2264}}
\email{suresh.kumar@plaksha.edu.in}
\affiliation{Data Science Institute, Plaksha University, Mohali, Punjab-140306, India}
\affiliation{Department of Mathematics, Indira Gandhi University, Meerpur, Haryana 122502, India}

\author{Rafael C. Nunes\,\orcidlink{0000-0002-8432-5616}}
\email{rafadcnunes@gmail.com}
\affiliation{Instituto de F\'{i}sica, Universidade Federal do Rio Grande do Sul, 91501-970 Porto Alegre RS, Brazil}
\affiliation{Divis\~ao de Astrof\'isica, Instituto Nacional de Pesquisas Espaciais, Avenida dos Astronautas 1758, S\~ao Jos\'e dos Campos, 12227-010, SP, Brazil}

\author{Emre \"{O}z\"{u}lker\,\orcidlink{0000-0003-0817-4219}}
\email{e.ozulker@sheffield.ac.uk}
\affiliation{Department of Physics, Istanbul Technical University, Maslak 34469 Istanbul, T\"{u}rkiye}
\affiliation{School of Mathematics and Statistics, University of Sheffield, Hounsfield Road, Sheffield S3 7RH, United Kingdom}

\author{J. Alberto Vazquez\,\orcidlink{0000-0002-7401-0864}}
\email{javazquez@icf.unam.mx}
\affiliation{Instituto de Ciencias F\'isicas, Universidad Nacional Aut\'onoma de M\'exico, Cuernavaca, Morelos, 62210, M\'exico}

\author{Anita Yadav\,\orcidlink{0009-0005-9868-8425}}
\email{anita.math.rs@igu.ac.in}
\affiliation{Department of Mathematics, Indira Gandhi University, Meerpur, Haryana 122502, India}

\date{\today}
\begin{abstract}
We have successfully integrated $\Lambda_{\rm s}$CDM, a promising scenario for alleviating major cosmological tensions, into a concrete theoretical framework by endowing it with a specific Lagrangian from the VCDM model, a type-II minimally modified gravity. This promotes the scenario to a fully predictive model (dubbed $\Lambda_{\rm s}$VCDM) that specifies the cosmological evolution self-consistently, including through the late-time AdS-to-dS transition epoch. In this theory, an auxiliary scalar field generates an effective cosmological constant in the Friedmann equation not only when endowed with a constant potential, but also when endowed with a linear potential. This property allows an abrupt mirror AdS-to-dS transition to be realised via a piecewise-linear potential, implemented as a sudden change in slope at a junction. To remove the associated sudden (type-II) singularity and ensure stable evolution, we smooth the junction using a blended sigmoid interpolant, obtaining rapid but continuous transitions. We identify two qualitatively distinct smooth mirror AdS-to-dS realisations of $\Lambda_{\rm s}$: (i) an \textit{agitated} transition, in which the potential interpolates between equal-magnitude AdS and dS plateaus and $\Lambda_{\rm s}$ generically develops a central bump; and (ii) a \textit{quiescent} transition, in which the potential remains continuous but changes slope across the transition layer, so that $\Lambda_{\rm s}(a)$ can remain monotone (possibly with shallow entrance/exit shoulders) and a central bump is not automatic. Depending on the transition type and sharpness, a finite-width transition can induce a transient accelerated-expansion interval ($\ddot a>0$) around the transition redshift ($z\sim 1.5$--$2$), in addition to the present-day accelerated expansion (for $z\lesssim0.6$ as in $\Lambda$CDM), and, if the background enters a region where $V_{,\phi\phi}>2/3$, a nested super-acceleration ($\dot H>0$) episode (and hence a bump in $H$). These distinct transient expansion histories can imprint characteristic signatures on both background and perturbation evolution; while the linear perturbation system is, in form, identical to that of $\Lambda$CDM, the scalar sector is modified through a $\dot H$-dependent relation, with deviations localised primarily to the transition epoch. Our construction therefore enables a self-consistent observational assessment of smooth $\Lambda_{\rm s}$CDM realisations and motivates dedicated multi-probe analyses to test transition dynamics and reassess cosmological tensions. Further work is warranted to assess whether $\Lambda_{\rm s}$CDM can emerge as a credible extension of the concordance model, or at least as a useful guide for exploring its potential revisions.
\end{abstract}
\maketitle
\section{Introduction}
The Hubble constant ($H_0$) tension is the foremost challenge in contemporary precision cosmology~\cite{Verde:2019ivm,DiValentino:2020zio,DiValentino:2021izs,Perivolaropoulos:2021jda,Schoneberg:2021qvd,Shah:2021onj,Abdalla:2022yfr,DiValentino:2022fjm,Kamionkowski:2022pkx,Giare:2023xoc,Hu:2023jqc,Verde:2023lmm,DiValentino:2024yew,Perivolaropoulos:2024yxv,CosmoVerseNetwork:2025alb}. Although the cosmological constant (CC) problem~\cite{Weinberg:1988cp,Peebles:2002gy} has a longer history, the $H_0$ tension seems more relevant to low-energy physics, a domain previously thought to be well understood. This may be signaling possible new physics beyond the standard cosmological model, i.e., $\Lambda$CDM, if not stemming from unidentified systematics. The persistence of the tension across various probes and over time diminishes the possibility of systematics or statistical flukes in the data,\footnote{Achieving agreement on the value of $H_0$ is not possible by simply disregarding one or a few late-time probes~\cite{Freedman:2020dne,Birrer:2020tax,Wu:2021jyk,Anderson:2023aga,Scolnic:2023mrv,Jones:2022mvo,Anand:2021sum,Freedman:2021ahq,Uddin:2023iob,Huang:2023frr,Li:2024yoe,Pesce:2020xfe,Kourkchi:2020iyz,Schombert:2020pxm,Blakeslee:2021rqi,deJaeger:2022lit,Murakami:2023xuy,Breuval:2024lsv,Freedman:2024eph,Riess:2024vfa,Vogl:2024bum,Gao:2025fcr,Scolnic:2024hbh,Said:2024pwm,Boubel:2024cqw,H0DN:2025lyy}, as the tension remains at $4$--$6\sigma$ even when accounting for different teams, objects, and calibrators~\cite{Riess:2019qba,DiValentino:2020vnx,DiValentino:2022fjm,H0DN:2025lyy}. For discussions regarding potential systematic effects in the data, see also~\cite{Dominguez:2019jqc,Park:2019emi,Lin:2019zdn,Boruah:2020fhl,Cao:2022ugh,Chen:2024gnu,Mortsell:2021nzg,Mortsell:2021tcx,Riess:2021jrx,Sharon:2023ioz,Murakami:2023xuy,Riess:2023bfx,Bhardwaj:2023mau,Brout:2023wol,Dwomoh:2023bro,Uddin:2023iob,Riess:2024ohe,Freedman:2024eph,Riess:2024vfa,H0DN:2025lyy}. Moreover, even if some late-time measurements do not exhibit strong tension with early universe probes due to large error bars, the puzzling characteristic remains: no late-universe measurements fall below the early ones, and vice versa, contrary to the expectation for measurements scattered around a true value~\cite{H0DN:2025lyy}.} yet, despite the plethora of attempts, there is no resolution through new physics that is both observationally and theoretically fully satisfactory or at least widely accepted. Moreover, addressing the $H_0$ tension while ensuring compatibility with all available data, without exacerbating other less definitive discrepancies such as the $S_8$ tension~\cite{DES:2021wwk,DiValentino:2020vvd,DiValentino:2018gcu,Kilo-DegreeSurvey:2023gfr,Troster:2019ean,Heymans:2020gsg,Dalal:2023olq,Chen:2024vvk,Kim:2024dmg,DES:2024oud,Harnois-Deraps:2024ucb,Dvornik:2022xap,Armijo:2024ujo,Lau:2024xrd,Qu:2024sfu,Adil:2023jtu,Akarsu:2024hsu}, remains a challenging task. See~\cite{DiValentino:2021izs,Perivolaropoulos:2021jda,Schoneberg:2021qvd,Shah:2021onj,Abdalla:2022yfr,DiValentino:2022fjm,Akarsu:2024qiq,CosmoVerseNetwork:2025alb} for recent reviews.

The $\Lambda_{\rm s}$CDM model~\cite{Akarsu:2019hmw,Akarsu:2021fol,Akarsu:2022typ,Akarsu:2023mfb} emerges as one of the promising models for addressing major cosmological tensions, viz., $H_0$, $M_{\rm B}$ (Type Ia Supernovae absolute magnitude), and $S_8$ (growth parameter) tensions, along with some other less significant tensions, and stands as the most economical model with this capability. As shown in a series of studies (e.g.\ \cite{Akarsu:2021fol,Akarsu:2022typ,Akarsu:2023mfb,Akarsu:2024eoo,Yadav:2024duq,Escamilla:2025imi}), raising $H_0$ to SH0ES values naturally suppresses the present-day growth rate $f_0\simeq \Omega_{\rm m0}^{\gamma}$ (while retaining the GR benchmark $\gamma\simeq 0.55$~\cite{Linder:2005in,Wang:1998gt} for the growth index) and the clustering metric $S_8\equiv \sigma_8\sqrt{\Omega_{\rm m0}/0.3}$ (even though $\sigma_8$ increases slightly). This allows $\Lambda_{\rm s}$CDM to simultaneously alleviate the $H_0$ and $S_8$ tensions and, at the same time, to ease the growth-index ($\gamma$) tension~\cite{Nguyen:2023fip,Specogna:2023nkq} by bringing the inferred growth behaviour closer to the GR benchmark, \emph{without} invoking modified gravity. Moreover, the framework remains compatible with eBOSS Ly$\alpha$ BAO data at $z_{\rm eff}\sim 2.3$~\cite{eBOSS:2019ytm,eBOSS:2019qwo}, with estimates of the present age of the Universe from the oldest globular clusters, and (with neutrino parameters treated as free~\cite{Yadav:2024duq}) with standard neutrino properties. Notably, transitions occurring near $z_\dagger \simeq 1.7$ have been found in some analyses to allow these effects to occur simultaneously, though its preference remains model- and dataset-dependent. It was inspired by a recent conjecture based on the findings on graduated dark energy (gDE)~\cite{Akarsu:2019hmw}, proposing that the Universe has, around redshift ${z_\dagger\sim 2}$, undergone a rapid \textit{mirror} anti-de Sitter (AdS) vacuum to a de Sitter (dS) vacuum transition (a \textit{mirror} AdS-to-dS transition, corresponding to a sign-switching CC while maintaining the magnitude before and after the transition), while leaving all other constituents of the standard cosmology---e.g., cold dark matter, baryons, and the inflationary paradigm---as in the standard $\Lambda$CDM model. From both mathematical and physical perspectives, $\Lambda_{\rm s}$CDM is identical to $\Lambda$CDM for $z<z_{\dagger}$, featuring a positive CC after the transition, but introduces modifications for $z>z_{\dagger}$, characterized by a negative CC prior to the transition, extending back to the early Universe, including the recombination era at $z_{\rm rec} \sim 1100$ and beyond. However, from a phenomenological perspective---i.e., in terms of the Universe’s expansion dynamics and observational signatures---the modifications are effectively confined to redshifts $z \lesssim z_{\dagger}$, with the free parameter $z_\dagger\sim2$ estimated through robust statistical analyses using cosmological data~\cite{Akarsu:2019hmw, Akarsu:2021fol, Akarsu:2022typ, Akarsu:2023mfb, Akarsu:2024eoo}. Specifically, $\Lambda_{\rm s}$CDM replicates the $H(z)$ of $\Lambda$CDM for $z < z_\dagger$ but with larger values compared to $\Lambda$CDM, introduces deformations in $H(z)$ around $z \sim z_\dagger$, and for higher redshifts ($z \gtrsim 3$) becomes nearly indistinguishable from $\Lambda$CDM. Consequently, from a phenomenological standpoint, $\Lambda_{\rm s}$CDM is best regarded as a post-recombination, or late-time, modification to $\Lambda$CDM. For a comprehensive explanation of how the $\Lambda_{\rm s}$CDM framework differs from and resembles the standard $\Lambda$CDM framework, see~\cref{sec:appC}. In what follows, we focus on the physical perspectives of this scenario, particularly the realization of the proposed AdS-to-dS transition. The suggested rapid nature of the sign-switching CC, along with its shift from negative to positive values, presents challenges in identifying a concrete physical mechanism; with that said, the phenomenological success of $\Lambda_{\rm s}$CDM despite its simplicity, strongly encourages the search for possible underlying physical mechanisms. Recently, string-inspired realizations of $\Lambda_{\rm s}$ have been proposed, it was shown in~\cite{Anchordoqui:2023woo,Anchordoqui:2024gfa,Anchordoqui:2024dqc,Soriano:2025gxd} that although the AdS swampland conjecture suggests that AdS-to-dS transition in the late universe seems unlikely (due to the arbitrarily large distance between AdS and dS vacua in moduli space), it can be realized through the Casimir forces of fields inhabiting the bulk (see also~\cite{Lehnert:2025izp}).~\footnote{It remains to be clarified whether the model in~\cite{Anchordoqui:2023woo} is complete, as it appears to lack a prescription for determining the perturbations.} In~\cite{Nyergesy:2025lyi}, it was shown that the modified thermal renormalization group study of asymptotically safe quantum gravity models at very high temperatures results in a negative CC while turns it into a positive parameter for low temperatures.   Furthermore, it was demonstrated in~\cite{Alexandre:2023nmh} that in various formulations of GR, it is possible to obtain a sign-switching CC through an overall sign change of the metric. A general-relativistic realization of the scenario has also been presented under the name Ph-$\Lambda_{\rm s}$CDM, in which a phantom scalar field with a hyperbolic-tangent potential drives a smooth mirror AdS-to-dS transition in its energy density~\cite{Akarsu:2025gwi,Akarsu:2025dmj}. In the teleparallel-gravity framework, the $f(T)$--$\Lambda_{\rm s}$CDM scenario provides an alternative smooth realization of $\Lambda_{\rm s}$CDM~\cite{Souza:2024qwd} (see also Ref.~\cite{Akarsu:2024nas}). We refer readers, without claiming to be exhaustive, to Refs.~\cite{Sahni:2002dx, Vazquez:2012ag, BOSS:2014hwf, Sahni:2014ooa, BOSS:2014hhw, DiValentino:2017rcr, Mortsell:2018mfj, Poulin:2018zxs, Banihashemi:2018oxo, Banihashemi:2018has, Akarsu:2019ygx, Li:2019yem, Visinelli:2019qqu, Ye:2020btb, Perez:2020cwa, Akarsu:2020yqa, Ruchika:2020avj, Calderon:2020hoc, Ye:2020oix, Paliathanasis:2020sfe, Acquaviva:2021jov, Bag:2021cqm, Sen:2021wld, Ozulker:2022slu, DiGennaro:2022ykp, Akarsu:2022lhx, Moshafi:2022mva, vandeVenn:2022gvl, Ong:2022wrs, Tiwari:2023jle, Malekjani:2023ple, Vazquez:2023kyx, Adil:2023ara, Paraskevas:2023itu, Wen:2023wes, DeFelice:2023bwq, Menci:2024rbq, Gomez-Valent:2024tdb, Wang:2024hwd, Tyagi:2024cqp, Toda:2024ncp, Dwivedi:2024okk, Gomez-Valent:2024ejh, Manoharan:2024thb, Pai:2024ydi, Mukherjee:2025myk, Giare:2025pzu, Keeley:2025stf, Efstratiou:2025xou, Silva:2025hxw, Scherer:2025esj, Wang:2025dtk, Bouhmadi-Lopez:2025ggl, Tamayo:2025xci, Bouhmadi-Lopez:2025spo, Hogas:2025ahb, Yadav:2025vpx, Pedrotti:2025ccw, Forconi:2025gwo} for further theoretical and observational studies---for model-independent and non-parametric reconstructions~\cite{Capozziello:2018jya,Wang:2018fng,Dutta:2018vmq,Bonilla:2020wbn,Bernardo:2021cxi,Escamilla:2021uoj,Bernardo:2022pyz,Gomez-Valent:2023uof,Medel-Esquivel:2023nov,Bousis:2024rnb,Colgain:2024ksa,Mukherjee:2025ytj,Gonzalez-Fuentes:2025lei,Tan:2025xas}---exploring DE models that yield negative energy densities, often consistent with a negative (AdS-like) cosmological constant, particularly for $z \geq 1.5-2$, and aimed at addressing major cosmological tensions.~\footnote{Phantom DE models---whose energy densities decrease with redshift and, like $\Lambda_{\rm s}$CDM, predict lower values than $\Lambda$CDM at high redshifts while typically staying positive, unlike in $\Lambda_{\rm s}$CDM---are well-known for alleviating the $H_0$ tension. Among these, the so-called \textit{phantom crossing model}~\cite{DiValentino:2020naf} (DMS20~\cite{Adil:2023exv,Specogna:2025guo}) stands out; a recent analysis~\cite{Adil:2023exv} reaffirming its success also revealed that its ability to assume negative densities for $z \gtrsim 2$---mimicking a AdS-like CC at higher redshifts---plays a central role in its effectiveness. Interacting DE (IDE) models~\cite{Kumar:2017dnp,DiValentino:2017iww,Yang:2018uae,Pan:2019gop,Kumar:2019wfs,DiValentino:2019jae,DiValentino:2019ffd,Lucca:2020zjb,Gomez-Valent:2020mqn,Kumar:2021eev,Nunes:2022bhn,Bernui:2023byc,Giare:2024smz,Sabogal:2025mkp} offer another avenue for addressing the $H_0$ tension, yet model-independent reconstructions of the IDE kernel~\cite{Escamilla:2023shf} do not preclude negative DE densities for $z \gtrsim 2$. Recent DESI BAO data---when analyzed using the CPL parametrization---provided more than $3\sigma$ evidence for dynamical DE~\cite{DESI:2024mwx}; however, a less noted finding is that non-parametric reconstructions of the DE density based on DESI BAO data also suggest the possibility of vanishing or negative} DE densities for $z \gtrsim 1.5-2$~\cite{DESI:2024aqx,Escamilla:2024ahl}, a phenomenon likewise observed in pre-DESI BAO data, particularly the SDSS BAO data~\cite{Escamilla:2023shf,Sabogal:2024qxs,Escamilla:2024ahl}.

In this \textit{paper}, we demonstrate that the $\Lambda_{\rm s}$CDM model~\cite{Akarsu:2019hmw,Akarsu:2021fol,Akarsu:2022typ,Akarsu:2023mfb} can be elevated to a theoretically complete physical cosmology, offering a fully predictive description of our universe within a type-II minimally modified gravity framework~\cite{DeFelice:2020eju, DeFelice:2020cpt, DeFelice:2020prd, DeFelice:2021xps, DeFelice:2022uxv, Jalali:2023wqh}.

\section{Rationale}
The simplest form of the $\Lambda_{\rm s}$CDM model~\cite{Akarsu:2021fol,Akarsu:2022typ,Akarsu:2023mfb} was constructed phenomenologically by replacing the CC in $\Lambda$CDM with an \textit{abrupt sign-switching CC} at redshift $z_\dagger$ (the only additional free parameter, compared to $\Lambda$CDM, and subject to observational constraints): ${{\Lambda\,\,\rightarrow\,\,\Lambda_{\rm s}\equiv\Lambda_{\rm s0}\,{\rm sgn}[z_\dagger-z]}}$, where the transition employs the signum function (sgn), and ${\Lambda_{\rm s0}>0}$ denotes the present-day value of $\Lambda_{\rm s}$. Accordingly, the Friedmann equation~\footnote{This study focuses exclusively on spatially flat and uniform cosmology; it can, however, be readily extended to non-flat scenarios.} can be written as $3\Mpl^2 H^2 = \sum_I\rho_I + \rho_{\Lambda_{\rm s0}}\, \mathrm{sgn}[a-a_\dagger]$, where $H$ is the Hubble parameter, $\rho_I$ represents the energy densities of each standard matter field, and ${\rho_{\Lambda_{\rm s0}}=\Lambda_{\rm s0}\Mpl^2}$ is the energy density of the present-day CC, with $a_\dagger$ being the scale factor, $a$, at the time of the transition. Given its remarkable simplicity, representing a minimal deviation from $\Lambda$CDM, and its significantly superior statistical fit to the available data compared to $\Lambda$CDM~\cite{Akarsu:2021fol, Akarsu:2022typ, Akarsu:2023mfb}, the model suggests that the CC might indeed have undergone a sign switch sometime after recombination. This implies the profound possibility that the Universe featured, until recently ($z\sim2$), an AdS vacuum phase---a theoretical sweet spot welcomed due to the AdS/CFT correspondence~\cite{Maldacena:1997re} and preferred by string theory and string-theory-motivated supergravities~\cite{Bousso:2000xa}.~\footnote{Evidently, an epoch of inflation (driven by an inflaton) in the very early universe remains necessary to generate primordial fluctuations. Here, however, we explore the possibility of a tiny, negative CC serving as a \textit{bare} CC at early times, whose effects become significant only in the late universe, e.g., for $z\lesssim3$. Its value is purely data-driven, inferred to provide a better fit to observations, as demonstrated in the accompanying observational follow-up. Note that this framework does not aim to resolve the notoriously challenging cosmological constant problem.} However, to advance $\Lambda_{\rm s}$CDM beyond a phenomenological model and establish it as a predictive one, it is necessary to integrate it with a concrete theoretical mechanism/model that explains the sign-switch in the CC, as well as to smooth out the abrupt behaviour implemented by the signum function, since the resulting discontinuity leads to a type II (sudden) singularity~\cite{Barrow:2004xh}.~\footnote{See~\cite{Paraskevas:2024ytz} for a demonstration of the presence of a type II singularity in abrupt $\Lambda_{\rm s}$CDM scenario and its minimal impact on the formation and evolution of cosmic bound structures, thereby preserving the model's viability in this context.} One of the primary aims of this paper is to eliminate this singularity and provide precise predictions about the impact of the transition on cosmological observables. However, smoothing out the transition introduces a new challenge: we require a model or mechanism capable of facilitating such a transition, with the potential to allow scenarios in which $H$ increases during the transition, i.e., large values of $|\dot{H}|/(N H^2)$ (equivalent to $R$, the Ricci scalar), where $N$, the lapse function, defines the nature of the time variable $t$. The smooth AdS-to-dS transition can impart a kick to $\dot{H}/(N H^2)$, which, depending on the rapidity of the transition, may induce a temporary super-acceleration phase ($\dot{H}/(N H^2)>0$) in the Universe during the transition, potentially exerting a substantial influence on the theory’s observables; therefore, quantifying this effect unequivocally is integral to defining the model’s features. Moreover, it is well-recognized that super-accelerating behaviour can readily give rise to ghost-like degrees of freedom and/or instabilities in perturbation dynamics.~\footnote{Within GR, super-accelerated expansion behaviour (${H>0}$ with $\dot{H}/(N H^2)>0$)---realized, for example, through the introduction of a phantom scalar field---cannot occur without encountering ghost instabilities and/or violations of the weak energy condition. However, as demonstrated  in~\cite{DeFelice:2020eju, DeFelice:2022uxv}, the VCDM model, which can embed $\Lambda_{\rm s}$CDM, avoids these pathologies by leveraging recent advances in field theory applied to cosmology, such as employing timelike scalar fields to define a preferred frame.} For this reason, incorporating such a dynamic and smooth component into a minimal theory of gravity becomes a natural course of action, specifically within the theory known as VCDM, which features only tensor modes in the gravity sector and possesses a predictive Lagrangian,~\footnote{The theory of VCDM is described in detail in Appendix~A. In brief, it implements constraints in a non-linear way that transform a would-be scalar-tensor theory into a minimal theory, i.e., one with only gravitational waves in the gravity sector.} and offers a smooth limit for an evolution that approaches (even exponentially fast) $\Lambda$CDM dynamics. For further exploration of these aspects and other topics related to VCDM, see Refs.~\cite{DeFelice:2020eju, DeFelice:2020cpt, DeFelice:2020prd, DeFelice:2021xps, DeFelice:2022uxv, Jalali:2023wqh}.

\section{Abrupt shifting cosmological constant and $\Lambda_{\rm s}$CDM}
\label{sec:abrupt}
As a preliminary step, before fully implementing the $\Lambda_{\rm s}$CDM scenario~\cite{Akarsu:2021fol, Akarsu:2022typ, Akarsu:2023mfb} within the VCDM framework~\cite{DeFelice:2020eju}---a goal we will ultimately achieve in the next section by adopting a smooth AdS-to-dS transition rather than an abrupt one---we first aim to illustrate how a shifting (sft) effective cosmological constant (CC) can, in principle, be realized through the auxiliary scalar field $\phi$ in VCDM, leveraging its unique properties. To this end, we begin by considering an idealized \textit{abrupt shifting CC} scenario, which, for example, can be parameterized simply as:
 \begin{equation}
    \Lambda_{\rm sft}(\phi) = \frac{\Lambda_{\rm sft,a} + \Lambda_{\rm sft,b}}{2} + \frac{\Lambda_{\rm sft,a} - \Lambda_{\rm sft,b}}{2} \, \mathrm{sgn}(\phi - \phi_{\rm c}),
    \label{eq:Lambda_signum}
\end{equation}
where the scalar field $\phi$ increases monotonically as the universe expands (${H>0}$), i.e., ${\frac{{\rm d}\phi}{{\rm d}t}=-H(1+z)\frac{{\rm d}\phi}{{\rm d}z}>0}$ (cf.~\cref{sec:appA}). Consequently, $\Lambda_{\rm sft}(\phi)=\Lambda_{\rm sft,b}(\phi)$ holds true for $\phi<\phi_{\rm c}$---i.e, before the transition at $\phi = \phi_{\rm c}$---while $\Lambda_{\rm sft}(\phi)=\Lambda_{\rm sft,a}(\phi)$ applies afterward ($\phi>\phi_{\rm c}$). By imposing a \textit{mirror} AdS-to-dS transition condition ($\Lambda_{\rm sft,a}=-\Lambda_{\rm sft,b}>0$), this construction reduces to an abrupt $\Lambda_{\rm s}$CDM scenario~\cite{Akarsu:2021fol,Akarsu:2022typ,Akarsu:2023mfb}:
\begin{equation}
\label{sgn:phi}
    \Lambda_{\rm s}(\phi) = \Lambda_{\rm s0} \, \mathrm{sgn}[\phi - \phi_{\rm c}],
\end{equation}
where $\Lambda_{\rm s0}\equiv \Lambda_{\rm s}(z=0)$ denotes the present-day value of the effective CC; for the mirror transition it coincides with the constant post-transition plateau, $\Lambda_{\rm s0}=\Lambda_{\rm sft,a}$.

\begin{figure*}
    \centering
    \includegraphics[width=0.45\textwidth]{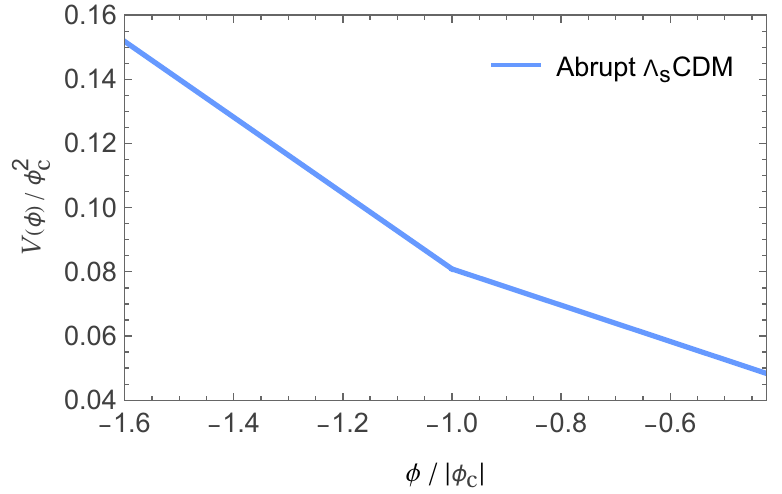}\hspace{1truecm}
    \includegraphics[width=0.45\textwidth]{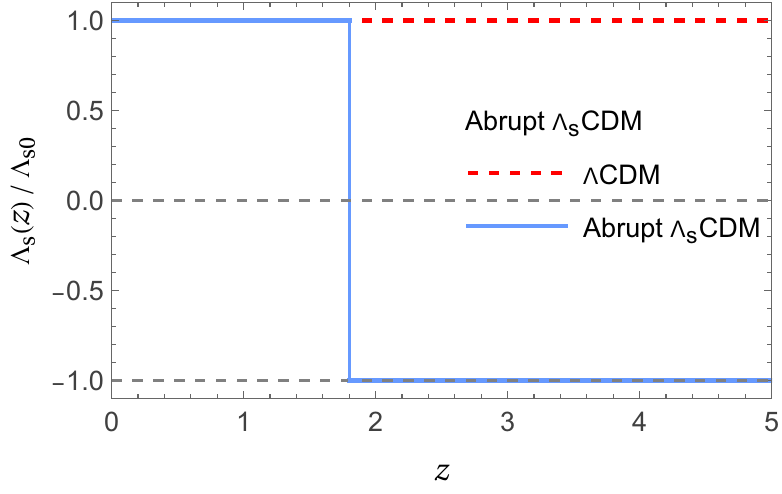}\vspace{0.5truecm}
    \includegraphics[width=0.45\textwidth]{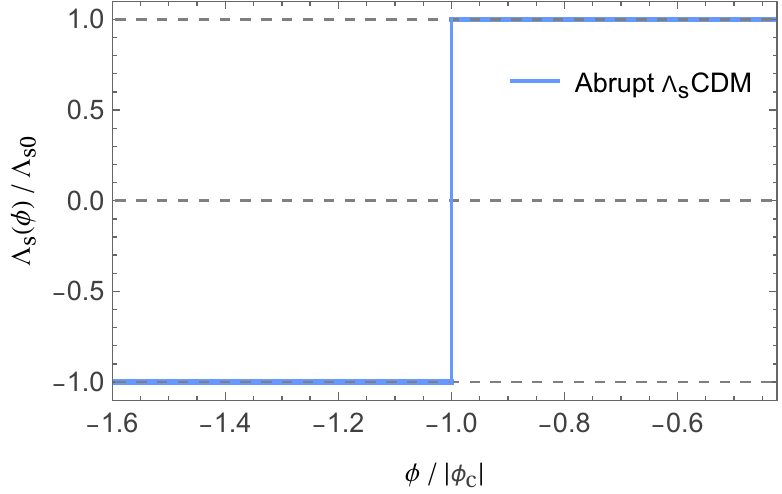}\hspace{1truecm}
    \includegraphics[width=0.45\textwidth]{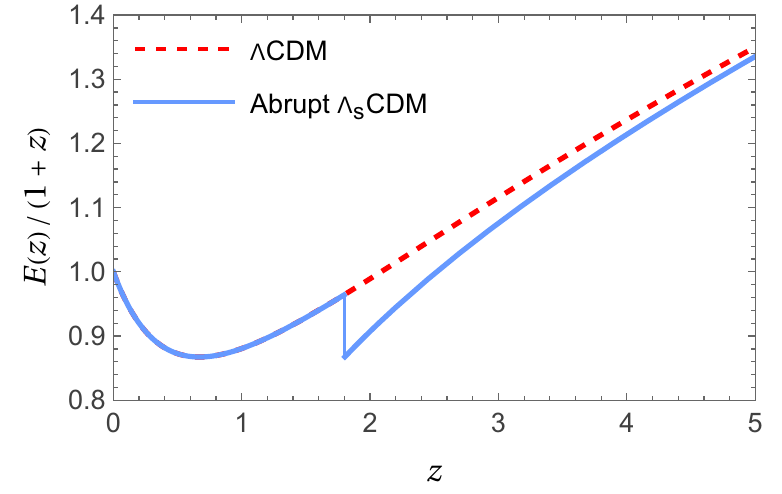}
    \caption{\textit{Abrupt (step) mirror AdS-to-dS transition.}
\textbf{Top left:} Potential $V(\phi)$ (scaled by $\phi_{\rm c}^2$) over $-1.6\!\le\!\phi/|\phi_{\rm c}|\!\le\!\phi_{0}/|\phi_{\rm c}|\approx-0.424$ (with $\phi_{\rm c}/H_0\approx -8.851$), showing a sharp kink: the potential is continuous but its slope jumps at the transition, located at $\phi/|\phi_{\rm c}|=-1$.
\textbf{Bottom left:} Corresponding normalised effective cosmological constant $\Lambda_{\rm s}(\phi)/\Lambda_{\rm s0}$, with $\Lambda_{\rm s0}\equiv\Lambda_{\rm s}(z=0)$, exhibiting an abrupt mirror AdS-to-dS transition---i.e., a discontinuous jump from $-\Lambda_{\rm s0}$ to $+\Lambda_{\rm s0}$ across $\phi_{\rm c}$.
\textbf{Top right:} Normalised effective cosmological constant $\Lambda_{\rm s}(z)/\Lambda_{\rm s0}$ (solid), compared with the $\Lambda$CDM baseline (red dashed $=1$); the abrupt transition is set at $z_\dagger=1.8$.
\textbf{Bottom right:} Corresponding expansion history, shown as the normalised conformal Hubble rate $E(z)/(1+z)$ (equivalently, $\dot a/H_0$) for the abrupt model (solid) and $\Lambda$CDM (red dashed) over redshift range $0\!\le\!z\!\le\!5$, with $E(z)=H(z)/H_0$; an abrupt jump, coincident with that in the effective CC, appears at $z_\dagger=1.8$.
We adopt $\phi_0=\phi(z=0)=-3H_0+\tfrac32\,\alpha_{\rm a}$ with $\alpha_{\rm a}/H_0=-0.5$, leading to $\alpha_{\rm b}/H_0\approx-1.046$, $\phi_{\rm c}/H_0\approx-8.851$, and $\beta/H_0^2\approx-6.337$ via the model constraints.
We assume $\Omega_{\rm m0}=0.30$, $\Omega_{\rm r0}=9\times10^{-5}$, and a spatially flat universe, implying $\Omega_{\Lambda_{\rm s0}}=1-\Omega_{\rm m0}-\Omega_{\rm r0}$ and hence $\Lambda_{\rm s0}=3H_0^2(1-\Omega_{\rm m0}-\Omega_{\rm r0})$. Because the potential is non-differentiable at the transition, this abrupt case cannot be embedded in the VCDM framework, which requires a smooth (differentiable) realisation.
}
    \label{fig:VL-step}
\end{figure*}

Remarkably, in VCDM, realizing such scenarios does not necessitate an abrupt shift in the scalar field potential $V(\phi)$ itself; such transitions remain achievable even while the scalar field is monotonically rolling down the potential. To demonstrate this, we now proceed with a continuous, piecewise linear potential $V(\phi)$ composed of two segments with different slopes ($\alpha=V_{,\phi}$) before and after the critical scalar field value $\phi=\phi_{\rm c}$.~\footnote{Refer to~\cref{sec:appA} for the rationale behind using continuously connected linear potential segments and for a discussion on the possibility of an abrupt shift in the effective CC through a sudden (discontinuous) change in the value of $V(\phi)$ at $\phi=\phi_{\rm c}$, rather than merely in its slope. This scenario is also addressed later in the paper. Additionally, the illustrative, non-differentiable potential employed here as an illustrative example will later be replaced with a smooth (infinitely differentiable) version.} Specifically, we define:
\begin{equation}
    V(\phi) = 
    \begin{cases}
        \alpha_{\rm b} (\phi-\phi_{\rm c})-\beta\,, & \text{for } \phi < \phi_{\rm c}\,, \\
        \alpha_{\rm a} (\phi-\phi_{\rm c})-\beta\,, & \text{for } \phi \geq \phi_{\rm c}\,,
    \end{cases}\label{eq:broken_pot}
\end{equation}
where $\beta$ is a constant offset. Substituting this potential into the field equations (see~\cref{sec:appA}), we obtain an abrupt shift in the effective CC, $\Lambda_{\rm sft} = \rho_\phi / \Mpl^2$, as follows:
\begin{equation}
    \Lambda_{\rm sft}(\phi) = 
    \begin{cases}
       \Lambda_{\rm sft,b}=\frac{3}{4}\alpha_{\rm b}^2-\alpha_{\rm b}\phi_{\rm c}-\beta\,, & \text{for } \phi < \phi_{\rm c}\,, \\
       \Lambda_{\rm sft,a}=\frac{3}{4}\alpha_{\rm a}^2-\alpha_{\rm a}\phi_{\rm c}-\beta\,, & \text{for } \phi \geq \phi_{\rm c}\,.
    \end{cases}
\end{equation}
In each segment, $\Lambda_{\rm sft}(\phi)$ depends quadratically on the slope, $\alpha$; $\Lambda_{\rm sft} = \frac{3}{4}\alpha^2 - \alpha\phi_{\rm c} - \beta$, representing an upward-opening parabola bounded from below. Its minimum value, $\Lambda_{\rm sft, \min} = -\frac{1}{3}\phi_{\rm c}^2 - \beta$, occurs at $\alpha_{\text{min}} = \frac{2}{3}\phi_{\rm c}$. Notably, the effective CC can be negative near this minimum if $\beta > -\frac{1}{3}\phi_{\rm c}^2$; otherwise, it remains positive. However, when $\Lambda_{\rm sft, \min} < 0$, for sufficiently large $|\alpha|$, the positive quadratic term dominates, ensuring that $\Lambda_{\rm sft}$ can still take positive values for $\alpha$ values far from $\alpha_{\text{min}}$. In particular, setting $\Lambda_{\rm sft} = 0$ yields two real roots: $\alpha_{1,2} = \frac{2}{3}\left( \phi_{\rm c} \pm \sqrt{ \phi_{\rm c}^2 + 3\beta } \right)$, with $\alpha_1 < \alpha_2$. Thus, $\Lambda_{\rm sft} < 0$ for $\alpha_1 < \alpha < \alpha_2$, and $\Lambda_{\rm sft} > 0$ outside this interval.

The key feature of this idealized setup is that the abrupt shift in $\Lambda_{\rm sft}$---from one constant value to another at $\phi = \phi_{\rm c}$---stems from a sudden change in the slope of the linear potential, $\Delta \alpha\equiv\alpha_{\rm a}-\alpha_{\rm b}$, rather than from a discontinuous shift in $V(\phi)$ itself, as might be commonly expected. This change in slope induces a corresponding abrupt shift in the effective CC, $\Delta\Lambda_{\rm sft}\equiv\Lambda_{\rm sft,a}-\Lambda_{\rm sft,b}$, given by:
\begin{equation}
\Delta \Lambda_{\rm sft} = \frac{3}{2}\left[H_{\rm a}(\phi_{\rm c})+H_{\rm b}(\phi_{\rm c})\right]\,\Delta\alpha,
\end{equation}
where $H_{\rm a}(\phi_{\rm c}) \equiv \lim_{\phi \to \phi_{\rm c}^+} H(\phi)$ and $H_{\rm b}(\phi_{\rm c}) \equiv \lim_{\phi \to \phi_{\rm c}^-} H(\phi)$ are the (positive) values of the Hubble parameter immediately before and after the transition, respectively. We are particularly interested in scenarios with $\Delta \Lambda_{\rm sft} > 0$, allowing an initially negative $\Lambda_{\rm sft}$ to become positive, corresponding to an AdS-to-dS transition in the effective CC. For each linear potential segment, we have the relation $\phi=\frac{3}{2}\alpha-3H$  (see~\cref{eq:phi_H} in~\cref{sec:appA}), implying $\alpha>\frac{2}{3}\phi$ for an expanding universe ($H>0$). Accordingly, as the scalar field continues to evolve monotonically in the positive $\phi$-direction, the sudden slope change at $\phi=\phi_{\rm c}$ simultaneously induces an abrupt shift in the Hubble parameter:
\begin{equation}
    \Delta H(\phi_{\rm c}) = \frac{\Delta\alpha}{2},
\end{equation}
where $\Delta H(\phi_{\rm c}) \equiv H_{\rm a}(\phi_{\rm c})-H_{\rm b}(\phi_{\rm c})$. Thus, $\Delta\alpha > 0$ leads to a positive shift in the expansion rate ($\Delta H(\phi_{\rm c}) > 0$), which can be leveraged to address the $H_0$ tension, while $\Delta\alpha < 0$ results in a negative shift ($\Delta H(\phi_{\rm c}) < 0$). Since $\dot{\phi} > 0$, ensuring the scalar field rolls down the potential, requires $\alpha_{\rm b} < 0$ and $\alpha_{\rm a} < 0$. Combining this with $\Delta\alpha > 0$ (necessary for $\Delta H(\phi_{\rm c}) > 0$) gives $\alpha_{\rm b} < \alpha_{\rm a} < 0$. To achieve a shift in $\Lambda_{\rm sft}$ from negative to positive values, we further need $\alpha_1 < \alpha_{\rm b} < \alpha_2$ (ensuring $\Lambda_{\rm sft,b} < 0$) and $\alpha_{\rm a} > \alpha_2$ (ensuring $\Lambda_{\rm sft,a} > 0$). Hence, it is possible to abruptly increase the effective CC---and consequently $H$---through a sudden flattening of the potential slope, i.e., $|\alpha_{\rm a}| < |\alpha_{\rm b}|$. An abrupt AdS-to-dS transition in the effective CC is also possible if $\Delta \Lambda_{\rm sft}$ is positive and sufficiently large to offset the initially negative $\Lambda_{\rm sft}$, as proposed by the $\Lambda_{\rm s}$CDM model. This model, considering a mirror AdS-to-dS transition with the additional condition $\Lambda_{\rm sft,a} = -\Lambda_{\rm sft,b} > 0$, has shown to be promising for addressing major cosmological tensions such as $H_0$ and $S_8$ discrepancies simultaneously~\cite{Akarsu:2021fol,Akarsu:2022typ,Akarsu:2023mfb}. In such cases, we require $\beta=\Lambda_{\rm s0}+\frac34\alpha_{\rm b}^2-\alpha_{\rm b}\phi_{\rm c}$ and the final slope to be:
\begin{equation}
    \alpha_{\rm a}=\frac{2 }{3}\phi_{\rm c} + \sqrt{\left(\frac{2 }{3}\phi_{\rm c}-\alpha_{\rm b}\right)^2+\frac{8 }{3}\Lambda_{\rm s0}}\,.
\end{equation}
Here, the plus sign in front of the second fraction is selected to ensure that $\phi(z=0)>\phi_{\rm c}$, which requires $\alpha_{\rm a}>\frac{2}{3}\phi_{\rm c}+2H_0$, as derived from~\cref{eq:phi_H}.
For a comparison with $\Lambda$CDM, an example of the abrupt (step) $\Lambda_{\rm s}$CDM transition is illustrated in~\cref{fig:VL-step}, which shows the piecewise-linear potential $V(\phi)$, the corresponding discontinuous effective cosmological constant $\Lambda_{\rm s}(\phi)$, its redshift evolution $\Lambda_{\rm s}(z)/\Lambda_{\rm s}(0)$, and the associated expansion history, namely the normalised conformal Hubble rate $E(z)/(1+z)$  (with $E(z)\equiv H(z)/H_0$, where $H(z)$ is the Hubble parameter and $H_0$ its present value). For numerical illustrations throughout this paper we adopt present-day density parameters $\Omega_{\rm m0}=0.30$ for pressureless matter and $\Omega_{\rm r0}=9\times 10^{-5}$ for radiation, assuming spatial flatness. Note that there is a redundancy of parameters in this prescription; for a given $\phi_{\rm c}$, exactly the same $\Lambda_{\rm s}(\phi)$ can be achieved by infinitely many pairs of, say, $\{\alpha_{\rm b},\beta\}$.

We have thus demonstrated that an abrupt shifting effective CC---and consequently an abrupt $\Lambda_{\rm s}$CDM model featuring a mirror AdS-to-dS transition as a particular case---can indeed be realized within the VCDM framework. However, despite accurately capturing the background dynamics over most redshift ranges, this abrupt version cannot be fully integrated into VCDM. The underlying reason is that a discontinuity in the derivative of the potential at $\phi_{\rm c}$ induces a discontinuity in $H$,  resulting in a singularity in $\dot H$ (a crucial term in perturbation theory, as shown in~\cref{sec:appPert}). To address this issue, it becomes necessary to smooth out the transition. Consequently, we will now investigate a completely smooth ($C^\infty$) $\Lambda_{\rm s}$CDM model.

\section{$\Lambda_{\rm s}$VCDM: Smooth $\Lambda_{\rm s}$CDM within VCDM}

In this section, we extend the above findings by incorporating sigmoid functions, enabling a smooth transition between two linear regimes: at this point, $\phi$ does acquire the meaning of the auxiliary field inside VCDM which gives rise to the dynamical dark energy component, and $\phi(z)$ has its own equation of motion, making possible the map $V(\phi)\!\to\!V\bigl(\phi(z)\bigr)$ and, equivalently, $\Lambda_{\rm s}(\phi)\!\to\!\Lambda_{\rm s}\bigl(\phi(z)\bigr)$, and conversely allowing the reconstruction of $V(\phi)$ from a prescribed $\Lambda_{\rm s}(z)$ trajectory (see~\cref{sec:appA}).

The sigmoid function, distinguished by its characteristic S--shaped curve, offers a gradual and continuous shift from one value to another. Here, we utilize a typical sigmoid function called the logistic function; it is expressed as:
\begin{equation}\label{eqn:logisticF}
    S(\phi) = \frac{1}{1 + e^{-2\eta(\phi - \phi_{\rm c})}},
\end{equation}
where $\eta>0$ controls the steepness (and hence the width) of the transition layer, and $\phi_{\rm c}$ is the centre (midpoint) of the interpolant, $S(\phi_{\rm c})=\tfrac12$, at which $|S'|$ is maximal; note for later reference that the zero-crossing of $\Lambda_{\rm s}(\phi)$ need not coincide with $\phi_{\rm c}$ (in general it occurs at some nearby $\phi_{\Lambda_{\rm s}=0}\neq\phi_{\rm c}$). It is often convenient to characterise the rapidity by the dimensionless parameter $\zeta\equiv\eta|\phi_{\rm c}|$, which naturally measures the transition sharpness in units of the field location: small $\zeta$ corresponds to broad/gentle blends, while large $\zeta$ signals sharper/narrower transitions. For late-time transitions of interest one typically has $|\phi_{\rm c}|/H_0=\zeta/(\eta H_0)=\mathcal{O}(1\text{--}10)$. For additional discussions on the rapidity/width of the transition epoch, see~\cref{sec:appA,sec:appFeature,app:bump}.

\subsection{Quiescent smooth transition}
\label{Qtrans}
To facilitate a \textit{quiescent smooth transition} (see~\cref{sec:appA} where this term is made precise) from the pre- to the post-transition linear regimes, we employ the sigmoid function as interpolant in the following blended potential, a formulation that is valid across all values of $\phi$:
\begin{equation}
    V = [\alpha_{\rm b} (\phi-\phi_{\rm c})-\beta][1 - S(\phi)] + [\alpha_{\rm a} (\phi-\phi_{\rm c})-\beta]S(\phi)\,,
    \label{eq:smooth_potential}
\end{equation}
with the corresponding effective CC
\begin{equation}
\begin{aligned}
\Lambda_{\rm s}(\phi)&=-\alpha_{\rm b}\phi_{\rm c}-\beta
-\Delta\alpha[\phi_{\rm c} S(\phi)+\phi(\phi-\phi_{\rm c})S'(\phi)]\\
&\quad+\frac{3}{4}[\alpha_{\rm b}+\Delta\alpha S(\phi)+\Delta\alpha (\phi-\phi_{\rm c})S'(\phi)]^{2},
\end{aligned}
\end{equation}
where one may substitute $S'(\phi)=2\eta\,S(\phi)\,[1-S(\phi)]$ when using the logistic interpolant $S(\phi)$ of~\cref{eqn:logisticF}. Equivalently, since
$S(\phi) = \tfrac{1}{2}+\tfrac{1}{2}\tanh[\eta(\phi-\phi_{\rm c})]$ (cf.~\cref{eqn:logisticF}), the potential~\eqref{eq:smooth_potential} can be cast in the compact form
\begin{equation}
    V(\phi)=\tfrac{\Sigma\alpha}{2} (\phi-\phi_{\rm c}) - \beta
    +\tfrac{\Delta\alpha}{2} (\phi-\phi_{\rm c})\tanh[\eta(\phi-\phi_{\rm c})],
    \label{eq:quiescent_tanh_potential}
\end{equation}
with the corresponding effective CC, $\Lambda_{\rm s}(\phi)=\rho_\phi/\Mpl^2$,
\begin{equation}
\begin{split}
\Lambda_{\rm s}(\phi)=&
-\frac{\Sigma\alpha}{2}\phi_{\rm c}-\beta
-\frac{\Delta\alpha}{2}\phi_{\rm c}
 \tanh[\eta(\phi-\phi_{\rm c})]
\\
&-\frac{\Delta\alpha}{2}\phi(\phi-\phi_{\rm c})\eta
 \sech^2[\eta(\phi-\phi_{\rm c})]
\\
&+\frac{3}{4}\biggl(
\frac{\Sigma\alpha}{2}
+\frac{\Delta\alpha}{2}\tanh[\eta(\phi-\phi_{\rm c})]
\\
&\quad\quad\,\,
+\frac{\Delta\alpha}{2}(\phi-\phi_{\rm c})\eta
 \sech^2[\eta(\phi-\phi_{\rm c})]
\biggr)^{2},
\end{split}
\label{eq:quiescent_tanh_potentialz}
\end{equation}
where $\Sigma\alpha\equiv\alpha_{\rm b}+\alpha_{\rm a}$. The potential introduced in~\cref{eq:smooth_potential} ensures a smooth ($C^\infty$) transition, with the interpolant function $S(\phi)$ increasing monotonically from $0$ to $1$ as $\phi$ traverses through the transition-layer centre $\phi_{\rm c}$. If one wishes to construct a smooth mirror AdS-to-dS transition---i.e., to connect the pre- and post-transition linear branches of $V(\phi)$ so that the effective CC evolves from $\Lambda_{\rm s}\approx -\Lambda_{\rm dS}$ to $\Lambda_{\rm s}\approx +\Lambda_{\rm dS}$, in accordance with the abrupt $\Lambda_{\rm s}$CDM limit---it is convenient to impose asymptotic (plateau) boundary conditions: far from the transition (equivalently, for $|\eta(\phi-\phi_{\rm c})|\gg 1$, so that $S\!\to\!0$ or $S\!\to\!1$), the effective CC saturates to the mirror values,
$\Lambda_{\rm s}(\phi)\to\Lambda_{\rm sft,b}=-\Lambda_{\rm dS}$ on the AdS (pre-transition) side and
$\Lambda_{\rm s}(\phi)\to\Lambda_{\rm sft,a}=+\Lambda_{\rm dS}$ on the dS (post-transition) side. Imposing these asymptotes fixes the parameters as
\begin{align}
    \beta&= \frac{3 \alpha_{\rm b}^{2} \alpha_{\rm a}+\left(4 \Lambda_{\rm dS}-3 \alpha_{\rm a}^{2}\right) \alpha_{\rm b}+4 \Lambda_{\rm dS} \alpha_{\rm a}}{4 \alpha_{\rm a}-4 \alpha_{\rm b}}\,,\\
    \phi_{\rm c}&=(8 \Lambda_{\rm dS}-3 \alpha_{\rm a}^{2}+3 \alpha_{\rm b}^{2})/(4 \alpha_{\rm b}-4 \alpha_{\rm a}),
\end{align}
with the consistency requirement ${\alpha_{\rm b} \neq \alpha_{\rm a}}$ (see~\cref{sec:appA}). At the transition-layer centre $\phi=\phi_{\rm c}$ one finds  
\begin{align}
   V(\phi_{\rm c}) &= -\beta,\\
   \Lambda_{\rm s}(\phi_{\rm c})
   &= -\beta - \tfrac{\Sigma\alpha}{2}\phi_{\rm c}
      + \tfrac{3}{16}\Sigma\alpha^{2}\\
    &= \tfrac{\Lambda_{\rm sft,a}+\Lambda_{\rm sft,b}}{2}
      - \tfrac{3}{16}(\Delta\alpha)^{2},\nonumber
\end{align}
In the mirror AdS-to-dS case ($\Lambda_{\rm sft,a}=-\Lambda_{\rm sft,b}=\Lambda_{\rm dS}>0$), this simplifies to  
\begin{equation}
   \Lambda_{\rm s}(\phi_{\rm c}) = -\tfrac{3}{16}(\Delta\alpha)^{2}\;\le\;0,
\end{equation}
so that the effective CC is necessarily non-positive at the transition-layer centre, while the potential remains $V(\phi_{\rm c})=-\beta$. 

\begin{figure*}
  \centering
  \includegraphics[width=0.45\textwidth]{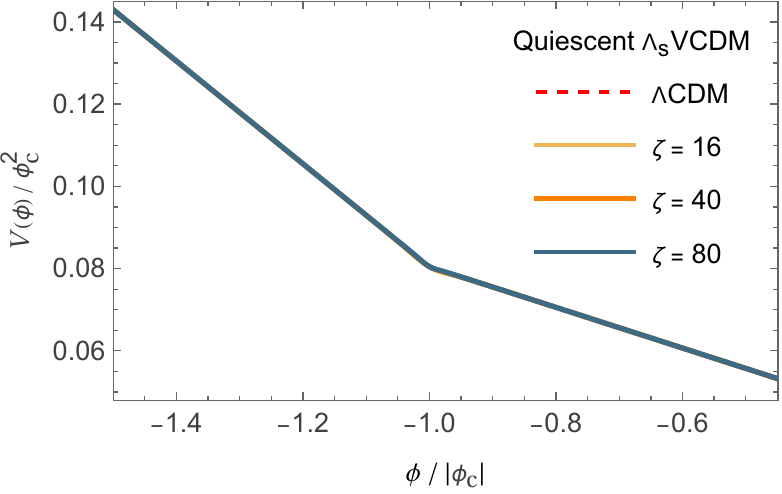}\hspace{1truecm}
  \includegraphics[width=0.45\textwidth]{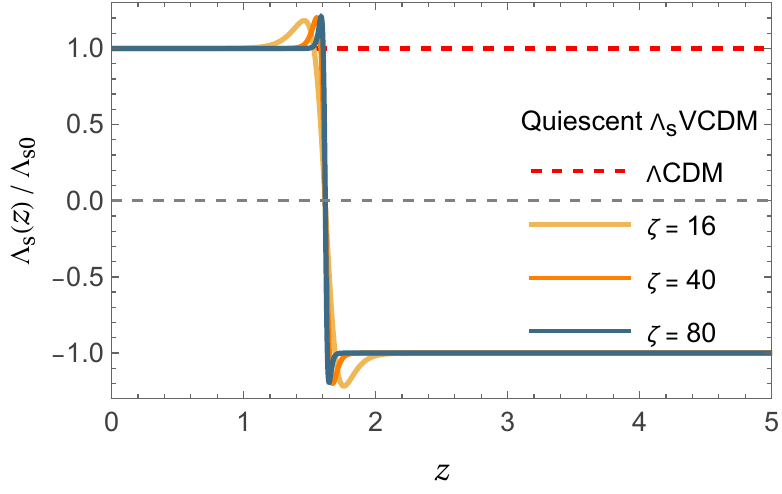}\vspace{0.5truecm}
  \includegraphics[width=0.45\textwidth]{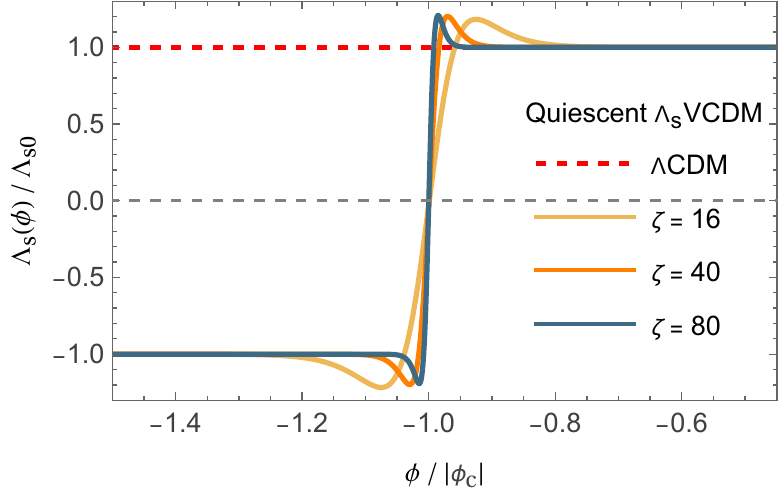}\hspace{1truecm}
  \includegraphics[width=0.45\textwidth]{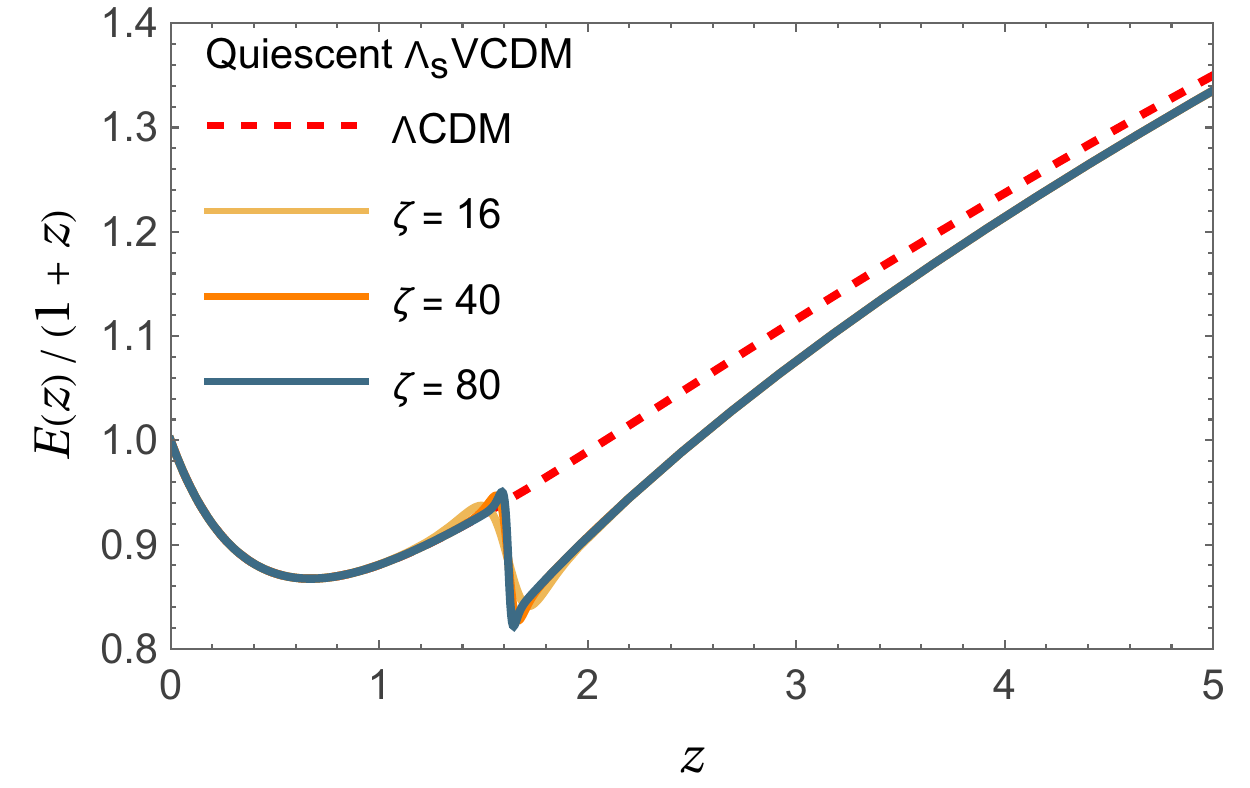}
  \caption{\textit{Quiescent mirror AdS-to-dS transition.}
\textbf{Top left:} Potential $V(\phi)$ (scaled by $\phi_{\rm c}^2$) for $\zeta=\eta|\phi_{\rm c}|=16,\,40,\,80$ (with the setting $\phi_{\rm c}/H_0=-8$), which set the rapidity/width of the transition, over $-1.5\!\le\!\phi/|\phi_{\rm c}|\!\le\!\phi_0/|\phi_{\rm c}|\approx-0.46$; the piecewise--linear potential’s slope changes smoothly and monotonically as the scalar field traverses the transition-layer centre at $\phi/|\phi_{\rm c}|=-1$.
\textbf{Bottom left:} Corresponding normalised effective cosmological constant $\Lambda_{\rm s}(\phi)/\Lambda_{\rm s0}$, with $\Lambda_{\rm s0}\equiv\Lambda_{\rm s}(z=0)$, showing a smooth mirror AdS-to-dS transition across $\phi_{\rm c}$, with localised non-monotonic ``shoulders'' near the entrance and exit of the transition; these arise where the curvature of the potential changes sign ($V_{,\phi\phi}=0$), consistent with ${\rm d}\Lambda_{\rm s}/{\rm d}\phi=3H\,V_{,\phi\phi}$.
\textbf{Top right:} Normalised effective cosmological constant $\Lambda_{\rm s}(z)/\Lambda_{\rm s0}$ over $0\!\le\!z\!\le\!5$ for the same $\zeta$ values, compared with the $\Lambda$CDM baseline (red dashed $=1$). \textbf{Bottom right:} Corresponding expansion history, shown as the normalised conformal Hubble rate $E(z)/(1+z)$ (equivalently, $\dot a/H_0$) for the three quiescent models (solid) and $\Lambda$CDM (red dashed) over $0\!\le\!z\!\le\!5$, with $E(z)=H(z)/H_0$; during the transition, $E(z)/(1+z)$ exhibits a local rise and shallow ``shoulders'' near the entrance and exit, co-located with those in $\Lambda_{\rm s}(z)$. In the quiescent examples shown, the universe undergoes accelerated expansion ($\ddot a>0$) not only today (for $z\lesssim0.6$, as in standard $\Lambda$CDM), but also over a short additional interval around $z\sim1.6$. A rise in $E(z)/(1+z)$ does not, by itself, necessarily imply a rise in the Hubble parameter $H(z)$ (not shown in the figure), i.e.\ a super–acceleration phase with $\dot H>0$; a genuine super–acceleration requires $V_{,\phi\phi}>2/3$ along the background, which for the logistic quiescent blend used here is equivalent to $\Delta\alpha\,\eta>2/3$. For the examples shown, this condition is satisfied and $H(z)$ develops a short super–acceleration phase---hence a bump---nested inside the transition. In all panels we adopt $\alpha_{\rm b}/H_0=-1$, $\phi_{\rm c}/H_0=-8$, $\Omega_{\rm m0}=0.30$, $\Omega_{\rm r0}=9\times10^{-5}$, and assume a spatially flat universe, implying $\Omega_{\Lambda_{\rm s0}}=1-\Omega_{\rm m0}-\Omega_{\rm r0}$ and hence $\Lambda_{\rm s0}=3H_0^2(1-\Omega_{\rm m0}-\Omega_{\rm r0})$. The function $E(z)=H(z)/H_0$ depends on $\Omega_{\rm m0}$, $\Omega_{\rm r0}$, and on the parameters of $\rho_{\phi}=\Mpl^2\Lambda_{\rm s}$ and the field $\phi$. The condition $E(0)=1$ can be solved for $\beta/H_0^2$ given the other parameters. We set the initial condition for the scalar field $\phi_0\equiv\phi(z=0)=-3H_0+\tfrac32\,\alpha_{\rm a}$. The condition $\Lambda_{\rm s}(\phi\ll\phi_{\rm c})=-\Lambda_{\rm s}(\phi_0)$ can be solved numerically for $\alpha_{\rm a}$ once $\eta H_0=\zeta H_0/|\phi_{\rm c}|$ is specified; for the values shown, $\alpha_{\rm a}/H_0\approx -0.396$. The field--redshift map is obtained by solving the equation of motion
${\rm d}(\phi/H_0)/{\rm d}z=-\tfrac{3}{2}\,[3\Omega_{\rm m0}(1+z)^3+4\Omega_{\rm r0}(1+z)^4]/[(1+z)E(z)]$.
}
  \label{fig:VL-sigmoid}
\end{figure*}

In the $\eta\!\to\!\infty$ limit, the smooth quiescent construction reduces to the abrupt model of~\cref{sec:abrupt}: the potential approaches the piecewise--linear form in~\cref{eq:broken_pot}, and the effective CC collapses to the step functions in~\cref{eq:Lambda_signum,sgn:phi}. For any finite $\eta$ the transition is $C^\infty$ and $\Lambda_{\rm s}(\phi)$ remains continuous; the instantaneous jump at $\phi=\phi_{\rm c}$ is recovered only in the strict $\eta\to+\infty$ limit (the value at $\phi_{\rm c}$ itself is conventional). Well before and after the transition---i.e., for $|\eta(\phi-\phi_{\rm c})|\gg1$ (practically, $\phi\lesssim\phi_{\rm c}-2/\eta$ and $\phi\gtrsim\phi_{\rm c}+2/\eta$)---one has $\Lambda_{\rm s}(\phi)\approx \Lambda_{\rm sft,b}$ on the entrance side and $\Lambda_{\rm s}(\phi)\approx \Lambda_{\rm sft,a}$ on the exit side. Along the realised background, $\phi(a)$ increases monotonically from values well below $\phi_{\rm c}$, traverses transition-layer centre $\phi=\phi_{\rm c}$, and approaches a finite late-time value $\phi_\infty=\tfrac{3}{2}\alpha_{\rm a}-3H_{\rm dS}$ with $H_{\rm dS}=\sqrt{\Lambda_{\rm dS}/3}$, corresponding to the asymptotic de Sitter future. Thus, the quiescent smooth transition interpolates continuously between the two constant plateaus of the abrupt model, with width $\Delta\phi \simeq 4\eta^{-1}$ (equivalently, in dimensionless form, $\Delta\phi/|\phi_{\rm c}| \simeq 4\zeta^{-1}$). Although $\Lambda_{\rm s}(\phi)$ follows a globally monotone envelope (i.e.\ the overall $\tanh$-driven trend) across the quiescent transition, localised departures from strict monotonicity can appear near the entrance and exit of the transition layer. These arise because $S(\phi)$ introduces regions where the curvature $V_{,\phi\phi}$ changes sign (see~\cref{sec:appFeature}); in such cases, \textit{along the background}, $\Lambda_{{\rm s},\phi}=3H\,V_{,\phi\phi}$ can vanish at isolated points within the window, creating shallow bump--dip ``shoulders'' superimposed on an otherwise monotone profile. For the logistic blend of linear branches used here one has
\begin{equation}
\label{eq:Vpp_logistic_exact}
\begin{aligned}
  V_{,\phi\phi}(\phi)
=
&\Delta\alpha\,\eta\,
\sech^{2}\!\big[\eta(\phi-\phi_{\rm c})\big]\\
&\times\Big[1-\eta(\phi-\phi_{\rm c})\tanh\!\big(\eta(\phi-\phi_{\rm c})\big)\Big].  
\end{aligned}
\end{equation}
This curvature changes sign at $|\phi-\phi_{\rm c}|\simeq 1.20\,\eta^{-1}$ (i.e., $|\phi-\phi_{\rm c}|/|\phi_{\rm c}|\simeq 1.20/\zeta$); hence, whenever the background trajectory spans the full transition window, shoulder features are \emph{inevitable}. Their amplitude scales with the slope contrast and the transition sharpness (cf.~Eq.~\eqref{eq:Vpp_logistic_exact}); schematically $|V_{,\phi\phi}|\propto |\Delta\alpha|\,\eta$, so broader/gentler blends ($\Delta\phi\sim\eta^{-1}$ large) or small $|\Delta\alpha|$ suppress the shoulders, while sharper blends or larger $|\Delta\alpha|$ enhance them. Importantly, these non--monotonic shoulders are a smoothing artefact---absent in the abrupt model---and can be parametrically small (and practically unobservable) in the broad/low--contrast regime. This does \emph{not} mean shoulders are generic to all quiescent constructions: in the reconstructed quiescent model (cf.~\cref{fig:VL-recon} and~\cref{sec:appFeature}) the combination in~\cref{eq:Vpp_generalS} keeps a definite sign along the realised trajectory, and $\Lambda_{\rm s}(a)$ remains strictly monotone with no shoulders.

Recalling $H=\tfrac12 V_{,\phi}-\tfrac{\phi}{3}$, one immediately sees that an expanding universe ($H>0$) implies $V_{,\phi}>\tfrac{2}{3}\phi$. For a linear potential ${V_{,\phi\phi}=0}$ with ${V_{,\phi}=\alpha}$ (constant), one has ${H=\tfrac{\alpha}{2}-\tfrac{\phi}{3}}$ and hence ${H>0}$ requires $\phi<\tfrac{3}{2}\alpha$. Accordingly, in the pre- and post-transition regimes of a quiescent transition one expects $H\simeq\tfrac{\alpha_{\rm b}}{2}-\tfrac{\phi}{3}$ and $H\simeq\tfrac{\alpha_{\rm a}}{2}-\tfrac{\phi}{3}$, respectively (see \cref{sec:appA} for parameter conditions); on these linear branches $V_{,\phi\phi}\simeq0$ and ${\rm d}H/{\rm d}\phi\simeq-\tfrac13$. Throughout the \emph{transition phase} one still has ${\rm d}H/{\rm d}\phi=\tfrac12 V_{,\phi\phi}-\tfrac13$. Hence, if $V_{,\phi\phi}(\phi)<\tfrac{2}{3}$ everywhere along the trajectory, then ${\rm d}H/{\rm d}\phi<0$ at all points and $H$ never increases; if instead $V_{,\phi\phi}$ exceeds the $2/3$ threshold over some sub-interval, then ${\rm d}H/{\rm d}\phi>0$ within that bounded window and $H$ exhibits a temporary interior rise, with the entry/exit turning points determined by $V_{,\phi\phi}=2/3$. For the logistic blend used here, $V_{,\phi\phi}$ attains its global maximum ${V_{,\phi\phi}^{\max}=\Delta\alpha\,\eta}$ at $\phi=\phi_{\rm c}$ (cf.~\cref{eq:Vpp_logistic_exact}); therefore $V_{,\phi\phi}>2/3$ somewhere \emph{iff} ${\Delta\alpha\,\eta>2/3}$ (with ${\Delta\alpha>0}$), in which case the ${\rm d}H/{\rm d}\phi>0$ window is symmetric about $\phi_{\rm c}$. With monotonic rolling ($\dot\phi>0$) this corresponds to a brief \textit{super-acceleration phase} ($\dot H>0$), i.e., a bump in $H$ function, nested within the overall transition---for detailed investigation on this, see~\cref{app:bump}. Finally, when $\Lambda_{\rm s}$ develops shallow ``shoulders'' near the entrance and exit of the transition---coinciding with the loci where $V_{,\phi\phi}=0$ (at $|\phi-\phi_{\rm c}|\simeq 1.20\,\eta^{-1}$ for the logistic case)---the Hubble rate inherits co-located, gentle, inflection-like features; at those points ${\rm d}H/{\rm d}\phi=-\tfrac{1}{3}$, so the sign change of $V_{,\phi\phi}$ modifies only the \emph{curvature} of $H(\phi)$ (not the sign of its slope), yielding mild, localised features in $H(z)$ near the beginning and end of the transition.

To determine the dynamics of our constructed physical system, it is necessary to specify an initial condition for $\phi$. Specifically, to find $H(z)$ and $\phi(z)$, we must integrate the following differential equation:
\begin{equation}
    \frac{{\rm d}\phi}{{\rm d}z}=-\frac{3}{2}\,\frac{\rho+P}{\Mpl^2 H (1+z)}\,,
\end{equation}
where $H=\frac{1}{2}\,V_{,\phi}-\frac{\phi}{3}$, and an initial condition for $\phi$ at a certain reference redshift is required (see~\cref{sec:appA}). Since the background evolution of each matter component $\rho_I$ (and $P_I$) is known as a function of redshift $z$, the combination $\rho+P$ is therefore a prescribed function of $z$.

We have thus demonstrated that a smooth $\Lambda_{\rm s}$CDM can be realized within the VCDM framework using a sigmoid-based transition of the potential function, which ensures the continuity of the function and its derivatives, making it ideally suited for, e.g., cosmological applications. As an example, \cref{fig:VL-sigmoid} illustrates a quiescent $\Lambda_{\rm s}$CDM model that can be regarded as a smoothed counterpart of the abrupt case in \cref{fig:VL-step}. It shows the smooth potential $V(\phi)$ constructed via the logistic interpolant~\cref{eqn:logisticF} and the blended form~\cref{eq:smooth_potential} (equivalently,~\cref{eq:quiescent_tanh_potential}); the corresponding $\Lambda_{\rm s}(\phi)$ (cf.~\cref{eq:quiescent_tanh_potentialz}); the redshift evolution $\Lambda_{\rm s}(z)$; and the associated expansion history, shown through the normalised conformal Hubble rate $E(z)/(1+z)$ (where $E(z)\equiv H(z)/H_0$ with $H(z)$ the Hubble function and $H_0$ the Hubble constant), directly compared with the $\Lambda$CDM baseline. Note that $E(z)/(1+z)=\dot a/H_0$; thus, an increase in $E(z)/(1+z)$ (towards lower redshift, i.e.\ forward in time) signals a phase of accelerated expansion, $\ddot a>0$. However, a rise in $E(z)/(1+z)$ does not, by itself, necessarily imply a rise in the Hubble parameter $H(z)$, i.e.\ a super--acceleration phase with $\dot H>0$.

\begin{figure*}
    \centering
    \includegraphics[width=0.45\textwidth]{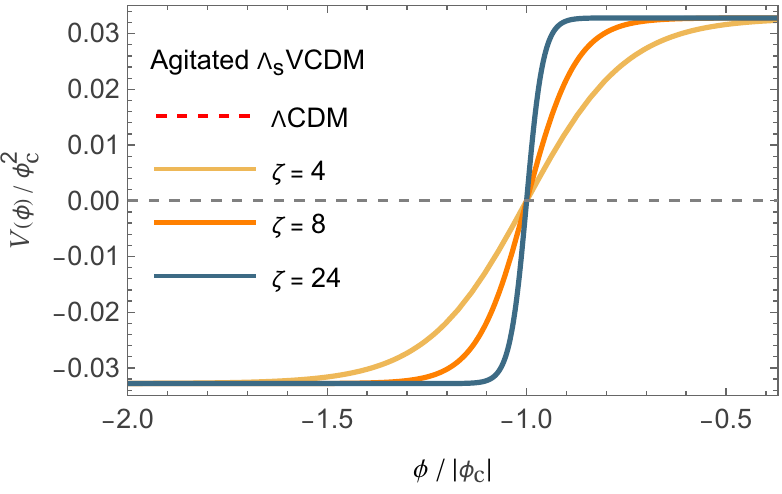}\hspace{1truecm}
  \includegraphics[width=0.45\textwidth]{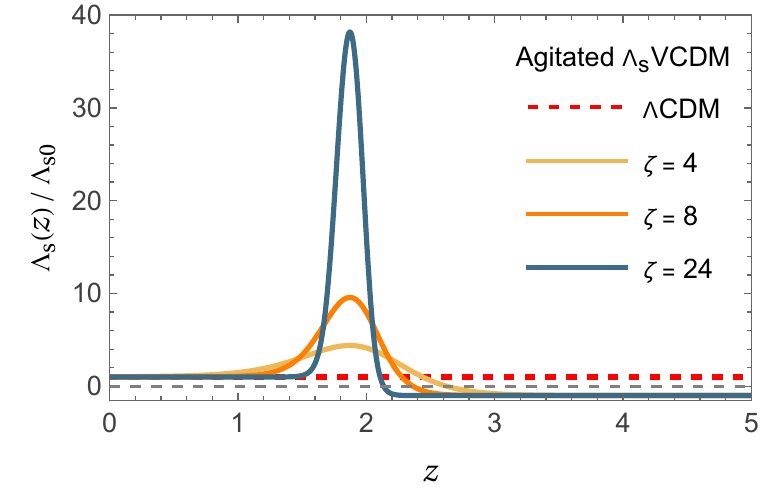}\vspace{0.5truecm}
  \includegraphics[width=0.45\textwidth]{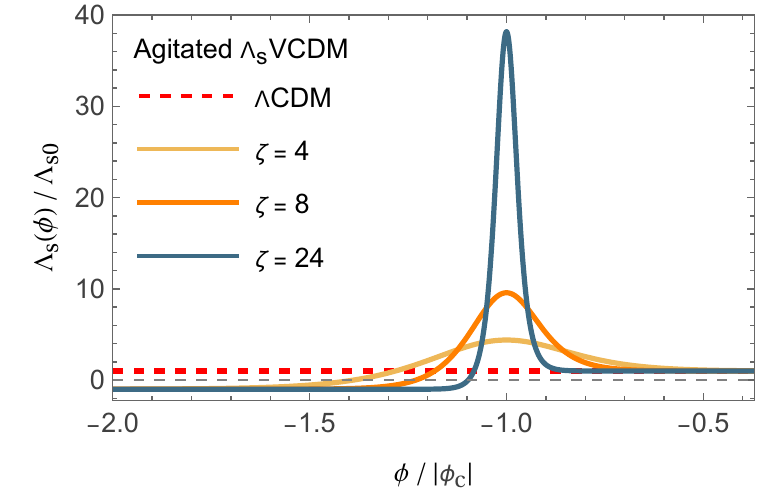}\hspace{1truecm}
  \includegraphics[width=0.45\textwidth]{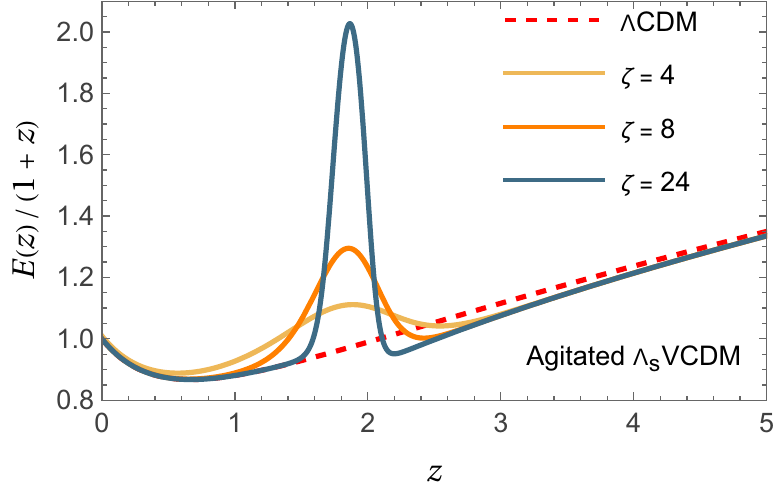}
   \caption{\textit{Agitated mirror AdS-to-dS transition.}
\textbf{Top left:} Potential $V(\phi)$ (scaled by $\phi_{\rm c}^2$) for $\zeta\equiv\eta|\phi_{\rm c}|=4,\,8,\,24$ (with the setting $\phi_{\rm c}/H_0=-8$), which control the rapidity/width of the transition, over $-2\!\le\!\phi/|\phi_{\rm c}|\!\le\!\phi_0/|\phi_{\rm c}|\approx-0.375$; the potential interpolates smoothly and monotonically from the AdS plateau to the dS plateau of equal magnitude as the scalar field passes through the transition-layer centre at $\phi/|\phi_{\rm c}|=-1$.
\textbf{Bottom left:} Corresponding normalised effective cosmological constant $\Lambda_{\rm s}(\phi)/\Lambda_{\rm s0}$, with $\Lambda_{\rm s0}\equiv\Lambda_{\rm s}(z=0)$, showing a localised bump centred at $\phi=\phi_{\rm c}$ that overshoots the dS plateau (expected if the transition has completed by today); the peak height increases and the width narrows as $\zeta$ increases (equivalently, as $\eta$ increases at fixed $\phi_{\rm c}=-8H_0$). The peak value is
$\Lambda_{\rm s}(\phi_{\rm c})=\zeta\,\Lambda_{\rm dS}+\tfrac{3}{4}\eta^{2}\Lambda_{\rm dS}^2$.
\textbf{Top right:} Normalised effective cosmological constant $\Lambda_{\rm s}(z)/\Lambda_{\rm s0}$ over the redshift range $0\!\le\!z\!\le\!5$ for the same $\zeta$ values, compared with the $\Lambda$CDM baseline (red dashed $=1$).
\textbf{Bottom right:} Corresponding expansion history, shown as the normalised conformal Hubble rate $E(z)/(1+z)$ (equivalently, $\dot a/H_0$) for the three agitated models (solid) and $\Lambda$CDM (red dashed) over $0\!\le\!z\!\le\!5$, with $E(z)=H(z)/H_0$; a bump in $E(z)/(1+z)$ appears, nearly coincident with that in $\Lambda_{\rm s}(z)$, whereas for sufficiently slow transitions the enhancement in $E(z)/(1+z)$ may remain gentle without an interior maximum. In the agitated examples shown, the universe undergoes accelerated expansion ($\ddot a>0$) not only today (for $z\lesssim0.6$, as in standard $\Lambda$CDM), but also over a short additional interval around $z\sim2.0$. A rise in $E(z)/(1+z)$ does not, by itself, necessarily imply a rise in the Hubble parameter $H(z)$, i.e.\ a super–acceleration phase with $\dot H>0$; a genuine super–acceleration requires $V_{,\phi\phi}>2/3$ along the background, which for the logistic agitated blend used here is equivalent to $\eta^{2}\Lambda_{\rm dS}>\sqrt{3}/2$. For the parameter choices in this figure, this condition translates to $\zeta\gtrsim5$; consequently, the $\zeta=8$ and $\zeta=24$ models develop a short super–acceleration phase---hence a bump---in $H(z)$ nested within the transition, whereas the $\zeta=4$ case remains in the gentle, non–bump regime. In all panels we adopt $\phi_{\rm c}/H_0=-8$, $\Omega_{\rm m0}=0.30$, $\Omega_{\rm r0}=9\times10^{-5}$, and assume a spatially flat universe, implying $\Omega_{\Lambda_{\rm s0}}=1-\Omega_{\rm m0}-\Omega_{\rm r0}$ and hence $\Lambda_{\rm s0}=3H_0^2(1-\Omega_{\rm m0}-\Omega_{\rm r0})$. The field--redshift map is obtained by solving the equation of motion
${\rm d}(\phi/H_0)/{\rm d}z=-\tfrac{3}{2}\,[3\Omega_{\rm m0}(1+z)^3+4\Omega_{\rm r0}(1+z)^4]/[(1+z)E(z)]$
with the initial condition $\phi_0\equiv\phi(z=0)=-3H_0$.} 
  \label{fig:VL_tanh}
\end{figure*}

\subsection{Agitated smooth transition}
\label{Atrans}
We now consider another type of smooth $\Lambda_{\rm s}$CDM scenario permitted within the VCDM framework: the \textit{agitated smooth transition} (see~\cref{sec:appA} where this term is made precise). Note that within VCDM, introducing a linear potential that suddenly changes slope is not the only method for achieving an abrupt shift in the effective CC. Another approach involves introducing a sudden shift from one constant potential value to another. Specifically, for an abrupt $\Lambda_{\rm s}$CDM limit, one could consider a potential abruptly jumping from $V_{\rm b}(\phi)=\text{const}<0$ to $V_{\rm a}(\phi)=-V_{\rm b}(\phi)$ at a critical point, viz., $\phi=\phi_{\rm c}$. To create a smooth version of this scenario, we again utilise the interpolant $S(\phi)$, but now blend the discontinuous \emph{constant} branch values directly, namely
\begin{equation}
    V(\phi)=V_{\rm b}\,[1-S(\phi)]+V_{\rm a}\,S(\phi)\,.
\end{equation}
Taking $V_{\rm a}=-V_{\rm b}\equiv\Lambda_{\rm dS}$ (i.e., a mirror offset about zero) then gives
\begin{equation}
V(\phi)=-\Lambda_{\rm dS}\,[1-S(\phi)]+\Lambda_{\rm dS}\,S(\phi).
\end{equation}
This construction can also be obtained by setting ${\alpha_{\rm b}=\alpha_{\rm a}=0}$ in~\cref{eq:gen_sigmoid_pot}. For the logistic interpolant $S(\phi)$ of~\cref{eqn:logisticF}, this yields the well-defined smooth potential
\begin{equation}
V(\phi)=\Lambda_{\rm dS}\tanh[\eta(\phi-\phi_{\rm c})],
\label{eq:tanh_potential}
\end{equation}
which itself realises a \emph{mirror} AdS–to–dS transition as $\phi$ traverses through the transition-layer centre $\phi_{\rm c}$: as $\phi$ increases monotonically from pre-transition times ($\phi\ll\phi_{\rm c}$) to post-transition times ($\phi\gg\phi_{\rm c}$), $V(\phi)$ evolves smoothly from the AdS plateau to the dS plateau of equal magnitude. This transition in the potential leads to the corresponding smoothly sign-switching effective CC:
\begin{equation}
\begin{aligned}
\Lambda_{\rm s}(\phi)
&= \Lambda_{\rm dS}\tanh[\eta(\phi-\phi_{\rm c})]
- \Lambda_{\rm dS}\,\eta\,\phi\,\sech^{2}[\eta(\phi-\phi_{\rm c})] \\
&\quad+ \frac{3}{4}\,\Lambda_{\rm dS}^{2}\eta^{2}\sech^{4}[\eta(\phi-\phi_{\rm c})].
\end{aligned}
\end{equation}
Because the transition has already completed by today ($z=0$)---i.e.\ $\phi_0\equiv\phi(z=0)>\phi_{\rm c}+2/\eta$ (completion criterion; see~\cref{sec:appA})---the field resides on the flat dS plateau, so $V(\phi_0)\simeq\Lambda_{\rm dS}$ and $\Lambda_{\rm s}(\phi_0)\simeq\Lambda_{\rm dS}$. Since $V_{,\phi}(0)\simeq0$, the kinematic relation $H=\tfrac12 V_{,\phi}-\tfrac{\phi}{3}$ (cf.~\cref{sec:appA}) implies $\phi_0\simeq-3H_0$. As $\phi$ increases monotonically with $a$ (cf.~\cref{sec:appA}), this requires $\phi_{\rm c}<\phi_0<0$. Equivalently, the completion criterion can be written as $|\phi_{\rm c}|>|\phi_0|+2/\eta$; multiplying both sides by $\eta>0$ yields $\zeta\equiv\eta|\phi_{\rm c}|>\eta|\phi_0|+2\ge2$. In what follows we therefore adopt $\zeta>2$, which guarantees that the AdS--to--dS transition has completed by today (with $\zeta\simeq2$ corresponding to a just-completed transition and $\zeta\gg2$ to one well in the past).

For the $\tanh$ potential in~\cref{eq:tanh_potential}, the curvature is
\begin{equation}
\label{agpotcurvature}
V_{,\phi\phi} \;=\; -\,2\,\eta^{2}\,\Lambda_{\rm dS}\,
\sech^{2}\!\big[\eta(\phi-\phi_{\rm c})\big]\,
\tanh\!\big[\eta(\phi-\phi_{\rm c})\big]\,,
\end{equation}
so $V_{,\phi\phi}$ changes sign \emph{once}, at $\phi=\phi_{\rm c}$. Using $\Lambda_{{\rm s},\phi}=3H\,V_{,\phi\phi}$ along the background (with $H>0$ and ${\rm d}\phi/{\rm d}a>0$), it follows that $\Lambda_{\rm s}(a)$ has a \emph{single} extremum at the centre: it rises for ${\phi<\phi_{\rm c}}$ and strictly falls for
${\phi>\phi_{\rm c}}$---a single, central bump (cf.~\cref{sec:appFeature}). For the logistic interpolant (cf.~\cref{eqn:logisticF}), the dimensionless curvature-to-slope ratio satisfies $|\phi_{\rm c}S''|/S'=2\,\zeta\,\big|\tanh[\eta(\phi-\phi_{\rm c})]\big|$, peaking at $2\zeta$; thus larger $\zeta$ yields a sharper, more pronounced central feature. From the explicit $\Lambda_{\rm s}(\phi)$ above, its value at the transition-layer centre is
\begin{equation}
\Lambda_{\rm s}(\phi_{\rm c}) \;=\; -\,\eta\,\Lambda_{\rm dS}\,\phi_{\rm c}
\;+\;\frac{3}{4}\,\eta^2 \Lambda_{\rm dS}^2 \;>\; 0
\quad(\phi_{\rm c}<0).
\end{equation}
Moreover, since $V_{,\phi\phi}<0$ for all $\phi>\phi_{\rm c}$, one has ${\rm d}\Lambda_{\rm s}/{\rm d}a<0$ after the centre while
$\Lambda_{\rm s}\!\to\!\Lambda_{\rm dS}$ as $a\!\to\!\infty$ (equivalently, $\phi\!\to\!\phi_{\infty}\equiv -3H_{\rm dS}=-\sqrt{3\Lambda_{\rm dS}}$); hence the late-time plateau is approached \emph{from above}, and the central value necessarily
exceeds the asymptote: $\Lambda_{\rm s}(\phi_{\rm c})>\Lambda_{\rm dS}$. In other words, an overshoot is a robust prediction of the agitated (matched--slope, offset--jump)
construction, and the peak occurs at the transition-layer centre, $\Lambda_{\rm s}^{\rm peak}=\Lambda_{\rm s}(\phi_{\rm c})$. The normalised peak height (peak--to--plateau ratio) can then be written as
\begin{equation}
\label{eqn:oscond}
\frac{\Lambda_{\rm s}^{\rm peak}}{\Lambda_{\rm dS}}
\;=\; \zeta \;+\; \frac{3}{4}\,\eta^{2}\,\Lambda_{\rm dS} \,>\, 1\,,
\end{equation}
where we impose $\zeta>2$ (completion criterion), which automatically implies the inequality. Sharper transitions (increasing $\zeta$ at fixed $\eta$, or increasing $\eta$ at fixed $\phi_{\rm c}$) produce larger overshoots. Overshoot only requires $\Lambda_{\rm s}^{\rm peak}/\Lambda_{\rm dS}>1$, i.e.\ $\zeta>1-\frac{3}{4}\eta^2\Lambda_{\rm dS}$; thus the completion criterion stated
above is a sufficient (and stronger) requirement, since it implies $\zeta>2$.

Having established that the $\tanh$ (agitated) realisation yields a single central bump in $\Lambda_{\rm s}$---and, provided the transition has completed by today, an overshoot above the late-time dS plateau---we now elaborate on the associated localised expansion dynamics. The same potential gives $V_{,\phi}=\eta\,\Lambda_{\rm dS}\sech^{2}[\eta(\phi-\phi_{\rm c})]$, which peaks at $\phi=\phi_{\rm c}$ with amplitude $\eta\Lambda_{\rm dS}$ and characteristic width $\sim\eta^{-1}$ (in field space). Via $H=\tfrac12 V_{,\phi}-\tfrac{\phi}{3}$, this narrow bump in $V_{,\phi}$ can produce a localised enhancement of $H$ (and $H^2$) near the transition redshift $z_\dagger$ as the field traverses $\phi_{\rm c}$, \emph{nearly coincident with} the overshoot of $\Lambda_{\rm s}$ driven by the same localised structure in $V$ and its derivatives. From~\cref{agpotcurvature} the positive maximum of the curvature is $V_{,\phi\phi}^{\max}=(4\sqrt{3}/9)\,\eta^{2}\Lambda_{\rm dS}$, attained at $\phi=\phi_{\rm c}-\eta^{-1}\tanh^{-1}\!\big(1/\sqrt{3}\big)<\phi_{\rm c}$; thus the strongest tendency to produce a local rise in $H$ (i.e.\ to maximise $H_{,\phi}$ and potentially allow $\dot H>0$) lies slightly on the AdS-side---within $\Delta\phi\simeq 0.658/\eta$ of the $\Lambda_{\rm s}$ peak at $\phi_{\rm c}$---and it is therefore natural to see a corresponding localised feature in $H(z)$ on the AdS side of the transition shortly after its onset; if sufficiently strong, this manifests as a genuine bump in $H(z)$: it increases over a short interval (i.e.\ $\dot H>0$) before resuming its overall decrease towards lower redshift. In particular, since $V_{,\phi\phi}>0$ only for $\phi<\phi_{\rm c}$ (cf.~\cref{agpotcurvature}) and $H_{,\phi}=\tfrac12 V_{,\phi\phi}-\tfrac13$ (equivalently, $\dot H=(\tfrac12 V_{,\phi\phi}-\tfrac13)\dot\phi$), any super-acceleration interval with $V_{,\phi\phi}>2/3$ (and hence $\dot H>0$ under our monotonic rolling, $\dot\phi>0$) is confined to a finite AdS-side window of the transition and exists \emph{iff} $\eta^{2}\Lambda_{\rm dS}>\sqrt{3}/2$. No super-acceleration occurs on the dS side of the transition, where $V_{,\phi\phi}<0$ and thus $H_{,\phi}<-\tfrac{1}{3}$ throughout. Note that the completion condition $\zeta>2$ guarantees an overshoot of $\Lambda_{\rm s}$, since it implies the overshoot criterion $\eta^{2}\Lambda_{\rm dS}>\frac{4}{3}(1-\zeta)$ (cf.~\cref{eqn:oscond}), but it does not by itself ensure super-acceleration; the latter requires the stronger bound $\eta^{2}\Lambda_{\rm dS}>\sqrt{3}/2$. When this bound holds, $H$ changes from decreasing to increasing at the AdS-side entry of the transition where $V_{,\phi\phi}=2/3$ (a local minimum), attains a local maximum at the AdS-side exit where $V_{,\phi\phi}$ returns to $2/3$ (both points with $\phi<\phi_{\rm c}$), and then resumes monotonically decreasing for $\phi\ge\phi_{\rm c}$---eventually relaxing to the late-time de Sitter asymptote $H_{\rm dS}=\sqrt{\Lambda_{\rm dS}/3}$ as $\phi\to\phi_{\infty}$ (or $t\to\infty$). See~\cref{app:bump} for further details. Thus, in the agitated scenario, the transition can lead to a brief \textit{super-acceleration phase} ($\dot H>0$), nested within the overall transition, occurring entirely on the AdS side shortly after its onset. Consistently, the dimensionless $|\dot{H}|/(N H^2)$ (cf.~\cref{eq:dot_H}) develops a nearly coincident feature controlled by $V_{,\phi\phi}\propto\eta^{2}\Lambda_{\rm dS}$, with the same width $\sim\eta^{-1}$. Since $\dot H$ enters directly into the only modified perturbation equation,~\cref{eqn:dotphi} (\cref{sec:appPert}), the transient propagates into linear perturbations. In short, $\eta$ controls the sharpness and strength of the signal---specifically, $\eta^{-1}$ sets the width in $\phi$-space (with peak amplitudes scaling as $V_{,\phi}(\phi_{\rm c})=\eta\,\Lambda_{\rm dS}$ and $V_{,\phi\phi}\sim\eta^{2}\Lambda_{\rm dS}$)---while the dimensionless $\zeta\equiv\eta|\phi_{\rm c}|$ controls the overall rapidity and, together with $\eta$, the overshoot of $\Lambda_{\rm s}$. These features of the agitated transitions therefore leave short-lived, potentially measurable imprints in both the background expansion and the evolution of perturbations.

As an example, \cref{fig:VL_tanh} shows the agitated case. The smooth potential $V(\phi)$ (\cref{eq:tanh_potential}) yields a $\Lambda_{\rm s}(\phi)$ profile with a \emph{single} central bump, in contrast to the shallow edge shoulders in the quiescent example of \cref{fig:VL-sigmoid}. The figure also displays the redshift evolution of $\Lambda_{\rm s}(z)$ and the normalised conformal Hubble rate, $E(z)/(1+z)$, overlaid with the $\Lambda$CDM baseline. The pronounced central bump in $\Lambda_{\rm s}$ around $\phi=\phi_{\rm c}$ is \emph{nearly coincident with} a localised enhancement in $H^2$ near the transition redshift $z_\dagger$. Thus, although the agitated and quiescent constructions share the same late-time effective CC, they exhibit distinct transient $H^2$ behaviour during the transition, providing a potential observational discriminant between the two scenarios at both the background and perturbative levels.

\section{Sigmoid $\Lambda_{\rm s}(a)$ and reconstruction of the scalar field potential}
\label{reconstLs}
\begin{figure*}
    \centering
      \includegraphics[width=0.45\textwidth]{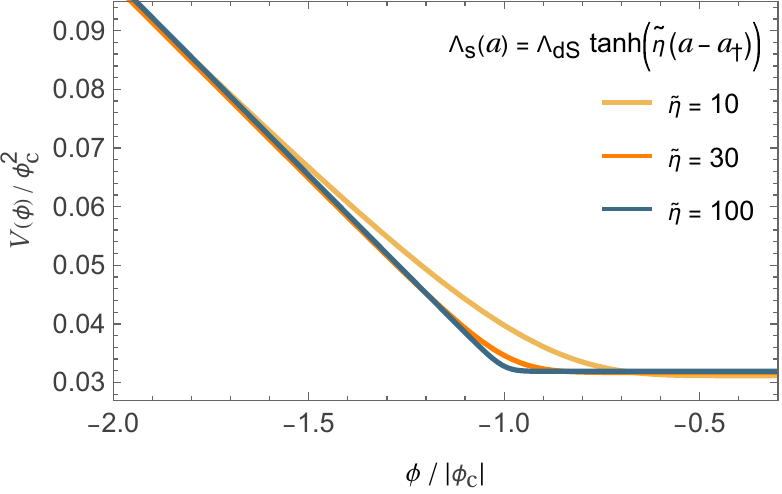}\hspace{0.5truecm}
       \includegraphics[width=0.45\textwidth]{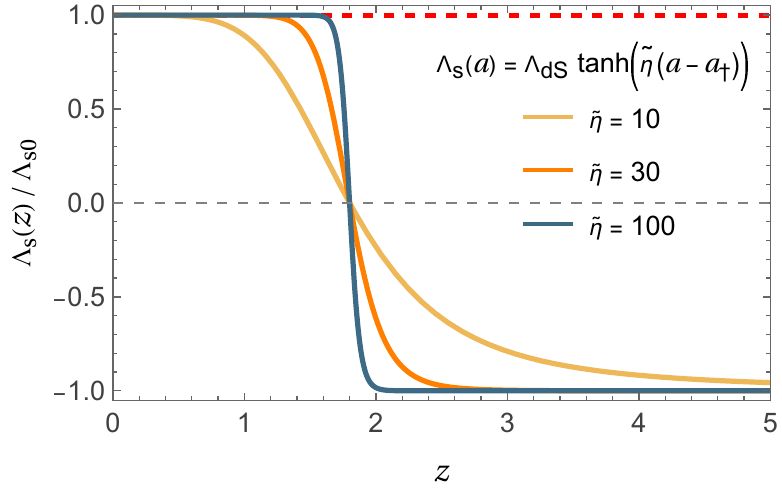}\vspace{0.5truecm}
       \includegraphics[width=0.45\textwidth]{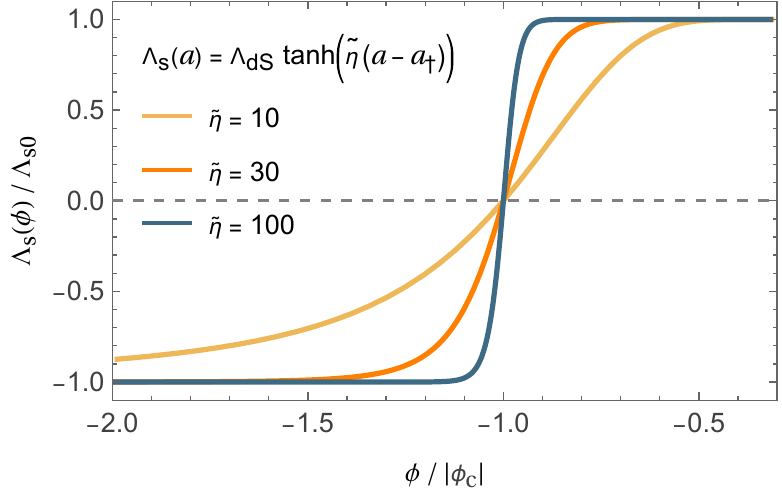}\hspace{0.5truecm}
      \includegraphics[width=0.45\textwidth]{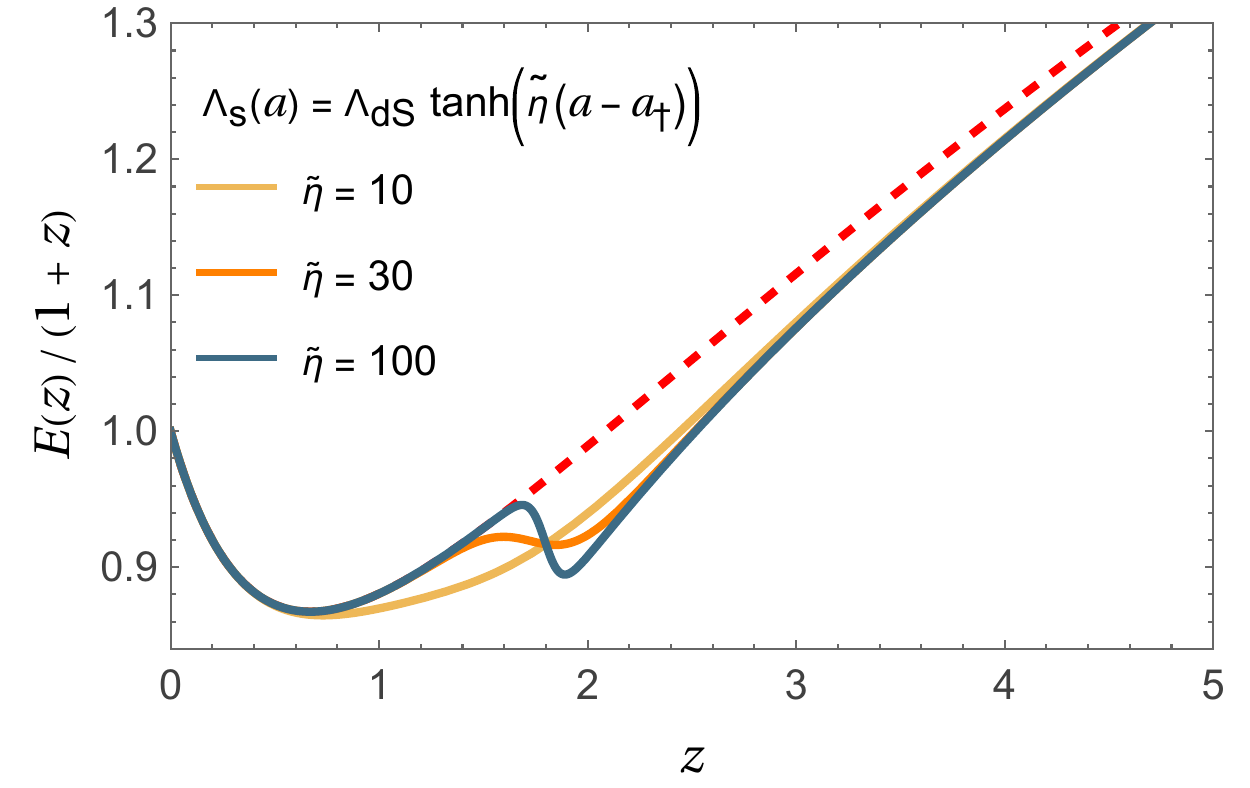} 
    \caption{\textit{Scale-factor-driven quiescent mirror AdS-to-dS transition.}
\textbf{Top left:} Potential $V(\phi)$ (scaled by $\phi_{\rm c}^2$) reconstructed from the imposed background solution for $\tilde\eta=10,\,30,\,100$ (yellow, orange, blue) over $-2\le\phi/|\phi_{\rm c}|\le-0.3$; the transition field values are $\phi_{\rm c}/H_0\simeq -8.21,\,-8.14,\,-8.11$, respectively. The reconstructed potential’s slope varies smoothly and monotonically as the field traverses the centre at $\phi/|\phi_{\rm c}|=-1$. The field--redshift map is obtained by solving
${\rm d}(\phi/H_0)/{\rm d}z=-\tfrac{3}{2}\,[3\Omega_{\rm m0}(1+z)^3+4\Omega_{\rm r0}(1+z)^4]/[(1+z)E(z)]$
with $E(z)=H(z)/H_0$ and the initial condition $\phi(z=0)=-3H_0$.
\textbf{Bottom left:} Corresponding normalised effective cosmological constant $\Lambda_{\rm s}(\phi)/\Lambda_{\rm s0}$, with $\Lambda_{\rm s0}\equiv\Lambda_{\rm s}(z=0)$, exhibiting a smooth, strictly monotonic sign switch from the negative to the positive asymptotic plateau---i.e., with no entrance/exit shoulders, unlike the quiescent example in \cref{fig:VL-sigmoid}.
\textbf{Top right:} Normalised effective cosmological constant $\Lambda_{\rm s}(z)/\Lambda_{\rm s0}$ (imposed; the other panels follow from this choice) over $0\!\le\!z\!\le\!5$ for the same $\tilde\eta$ values, compared with the $\Lambda$CDM baseline (red dashed $=1$). \textbf{Bottom right:} Corresponding expansion history, shown as the normalised conformal Hubble rate $E(z)/(1+z)$ (equivalently, $\dot a/H_0$) for the same $\tilde\eta$ values (solid) and $\Lambda$CDM (red dashed) over $0\!\le\!z\!\le\!5$, with $E(z)=H(z)/H_0$. In all these reconstructed examples the universe undergoes accelerated expansion ($\ddot a>0$) today (for $z\lesssim0.6$, as in standard $\Lambda$CDM). Adopting the fiducial values $\Omega_{\rm m0}=0.30$ and $z_\dagger\simeq1.8$, and using the analytic thresholds derived in \cref{reconstLs}, one finds that an additional accelerated–expansion interval (a local rise in $E(z)/(1+z)$) around the transition appears for $\tilde\eta\gtrsim\tilde\eta_{\rm acc}\simeq26$, whereas a super–acceleration regime ($\dot H>0$), and hence a genuine bump in $H(z)$ (not shown), occurs only for $\tilde\eta\gtrsim\tilde\eta_{\rm super}\simeq79$. Accordingly, the slowest case ($\tilde\eta=10<\tilde\eta_{\rm acc}$) does not develop an extra accelerated–expansion phase near $z\sim1.8$ and the transition appears only as a broad, mild flattening with respect to $\Lambda$CDM over $z\simeq1$–$2$. The intermediate case ($\tilde\eta=30>\tilde\eta_{\rm acc}$ but $<\tilde\eta_{\rm super}$) exhibits such an additional accelerated interval without entering a super–acceleration regime, while the sharpest case ($\tilde\eta=100>\tilde\eta_{\rm super}$) develops a brief super–acceleration phase nested inside the transition. All curves converge to the $\Lambda$CDM baseline at high redshift. Unlike in~\cref{fig:VL-step,fig:VL-sigmoid,fig:VL_tanh}, here the transition-layer centre is imposed at a fixed scale factor $a_\dagger$ (such that $z_\dagger\approx1.8$), so the corresponding $\phi_{\rm c}=\phi(z_\dagger)$ depends on the steepness/rapidity $\tilde{\eta}$ of the transition.}
    \label{fig:VL-recon}
\end{figure*}

We now explore an alternative approach to constructing a smooth $\Lambda_{\rm s}$CDM. Instead of deriving $\Lambda_{\rm s}(\phi)$ from a potential, we directly implement a smooth one as a function of scale factor $a$ (or redshift $z$), equivalently specifying the Hubble function $H(a)$, given that the behaviour of the matter sector is established. With the matter sector fixed, the dynamics of $\rho_\phi(a)$ are also automatically defined. Knowing $H(a)$, which yields the desired $\Lambda_{\rm s}(a)$, enables us to immediately determine the background and coefficients of the perturbation equations~\eqref{eqn:dotphi} without the need for integration. For instance, by utilising a well-known smooth approximation of the signum function (${\rm sgn}\,x\approx\tanh kx$, for constant $k>1$), $\Lambda_{\rm s}\equiv\Lambda_{\rm s0}\,{\rm sgn}[a-a_\dagger]$, we define (with scaling):
\begin{equation}\label{eqn:smoothLs}
   \Lambda_{\rm s}(a) = \Lambda_{\rm dS}\,\tanh\!\qty[\tilde\eta\,(a-a_{\dagger})],
\end{equation}
where $a_\dagger$ denotes the scale factor at the transition-layer centre (for the cases of interest $a_\dagger<1$), and we normalise $a$ so that $a=1$ today ($z=0$); the dimensionless parameter $\tilde\eta>1$ sets the rapidity/width of the transition (the width in scale factor scales as $\Delta a\sim 1/\tilde\eta$; full width at half maximum, FWHM $\simeq 1.763/\tilde\eta$), and $\Lambda_{\rm dS}=\Lambda_{\rm s0}/\tanh[\tilde\eta(1-a_\dagger)]$ ensures $\Lambda_{\rm s}(1)=\Lambda_{\rm s0}$. For a fast transition, e.g., $\tilde\eta\gtrsim10$, centred at $z_\dagger\simeq1.8$, we may safely set $\Lambda_{\rm dS}\approx\Lambda_{\rm s0}$, since $\tanh[\tilde\eta(1-a_\dagger)]\approx 1$.
A worked example is shown in \cref{fig:VL-recon}: starting from the sigmoid $\Lambda_{\rm s}(a)$ of~\cref{eqn:smoothLs}, we reconstruct the field history $\phi(z)$ and the potential $V(\phi)$; the figure also displays $\Lambda_{\rm s}(a)$ and the normalised conformal Hubble rate $E(z)/(1+z)$, overlaid with the $\Lambda$CDM baseline, for several choices of the rapidity $\tilde\eta$ at fixed $z_\dagger$, illustrating how the transition width $\propto \tilde\eta^{-1}$ maps into the background dynamics. 

For analytic control it is useful to characterise when the sigmoid $\Lambda_{\rm s}(a)$ in \cref{eqn:smoothLs} produces accelerated and super–accelerated expansion. Since the mirror AdS–to–dS transition occurs at late times ($z_\dagger=\mathcal{O}(1)$), we may safely neglect radiation for these estimates and work with a spatially flat matter+$\Lambda_{\rm s}$ background, expressed in terms of the normalised Hubble parameter $E^2(a)\equiv H^2(a)/H_0^2$,
\begin{equation}
\begin{aligned}
E^2(a)&=\Omega_{\rm m0}\,a^{-3}
+(1-\Omega_{\rm m0})\,F(a),\\
F(a)&=\frac{\tanh[\tilde\eta\,(a-a_\dagger)]}{\tanh[\tilde\eta\,(1-a_\dagger)]}.
\end{aligned}
\end{equation}
Using $\dot a=aH$ and $E=H/H_0$ one finds
\begin{equation}
\begin{aligned}
    \dot H&=\frac{aH_0^2}{2}\,\frac{{\rm d}E^2}{{\rm d}a},\\
\frac{\ddot a}{a}&=H^2+\dot H
=H_0^2\!\left(E^2+\frac{a}{2}\frac{{\rm d}E^2}{{\rm d}a}\right),
\end{aligned}
\end{equation}
where
\begin{equation}
\frac{{\rm d}E^2}{{\rm d}a}
=-3\,\Omega_{\rm m0}\,a^{-4}
+(1-\Omega_{\rm m0})\,\tilde\eta\,
\frac{\sech^2[\tilde\eta(a-a_\dagger)]}{\tanh[\tilde\eta(1-a_\dagger)]}.
\end{equation}
Thus an \emph{accelerated} phase ($\ddot a>0$) requires
\begin{equation}
2F(a)+a\,F_{,a}
>\frac{\Omega_{\rm m0}}{1-\Omega_{\rm m0}}\,a^{-3},
\end{equation}
while a \emph{super–acceleration} phase ($\dot H>0$) requires ${\rm d}E^2/{\rm d}a>0$, i.e.
\begin{equation}
F_{,a}>3\,\frac{\Omega_{\rm m0}}{1-\Omega_{\rm m0}}\,a^{-4},
\end{equation}
where the subscript $,_a$ denotes ${\rm d}/{\rm d}a$. The departure from the standard matter+$\Lambda$ background is encoded entirely in $F(a)$ and $F_{,a}$ and is localised around the transition-layer centre by the $\sech^2[\tilde\eta(a-a_\dagger)]$ factor, so the onset of any additional accelerated or super–accelerated interval is well captured by evaluating these inequalities near $a\simeq a_\dagger$. In the observationally relevant fast–transition regime $\tilde\eta(1-a_\dagger)\gg1$ (so that $\tanh[\tilde\eta(1-a_\dagger)]\simeq1$), one has at $a=a_\dagger$ the exact relations $F(a_\dagger)=0$ and $F_{,a}(a_\dagger)\simeq\tilde\eta$, which yield the approximate thresholds
\begin{equation}
\tilde\eta_{\rm acc}\;\simeq\;\frac{\Omega_{\rm m0}}{1-\Omega_{\rm m0}}\,a_\dagger^{-4},
\quad
\tilde\eta_{\rm super}\;\simeq\;3\,\tilde\eta_{\rm acc}.
\end{equation}
For the fiducial values $\Omega_{\rm m0}=0.30$ and $z_\dagger\simeq1.8$ ($a_\dagger\simeq0.357$) adopted in \cref{fig:VL-recon}, this gives $\tilde\eta_{\rm acc}\simeq26$ and $\tilde\eta_{\rm super}\simeq79$. Accordingly, the intermediate case $\tilde\eta=30$ lies above the acceleration threshold but below the super–acceleration threshold and therefore develops an additional accelerated–expansion interval without a genuine super–acceleration phase, whereas only the sharpest case $\tilde\eta=100$ crosses into a brief super–acceleration regime. This analytic expectation is in agreement with the behaviour seen in the bottom–right panel of \cref{fig:VL-recon}.

In this approach, we can straightforwardly further modify $\Lambda_{\rm s}(a)$, as given in~\cref{eqn:smoothLs}, to incorporate additional features into the $\Lambda_{\rm s}$CDM model at various redshifts, beyond the smooth mirror AdS-to-dS transition around $z_\dagger\sim2$ suggested by robust observational analysis (see, e.g., ~\cite{Akarsu:2019hmw, Akarsu:2021fol, Akarsu:2022typ, Akarsu:2023mfb, Akarsu:2024eoo}). However, we will leave the exploration of this possibility to future work. Having uniquely defined the $\Lambda_{\rm s}(a)$ function, we can now determine $H^2=\sum_I \varrho_I + \varrho_{\rm s}(a)$ (where the index $I$ runs over the standard matter components, and $\varrho_{\rm s}\equiv\Lambda_{\rm s}/3$ represents the effective CC contribution), and consequently also the expressions for $\dot{H}$ [see~\cref{eq:dotH}]. Hence, with $\Lambda_{\rm s}(a)$ specified, we have a complete model ready for direct implementation into a Boltzmann code (see~\cref{sec:appPert}). We emphasize that, aside from the additional free parameters $z_{\dagger}$ (or $a_\dagger$), which determine the redshift (or the scale) of the sign change in $\Lambda_{\rm s}$, and $\tilde\eta$, which governs the rapidity/width of the transition epoch, the entire parameter baseline of the $\Lambda_{\rm s}$CDM model---including cosmological initial conditions ($A_s$ and $n_s$), primarily set by the inflationary epoch---remains identical to that of the $\Lambda$CDM model. Furthermore, for scenarios focused exclusively on fast transitions, where data cannot distinguish between different transition rates, $\tilde\eta$ can be fixed to a sufficiently large value, effectively reducing the model to having only one additional free parameter, $z_{\dagger}$. These additional parameters ultimately govern the late-time modifications in the background dynamics---specifically in $H$ and $\dot{H}$---as well as in perturbations for $z\lesssim z_\dagger$ (with $z_\dagger\sim 2$ suggested by multiple observational analyses of abrupt $\Lambda_{\rm s}$CDM; see, e.g., \cite{Akarsu:2019hmw,Akarsu:2021fol,Akarsu:2022typ,Akarsu:2023mfb,Akarsu:2024eoo}), whereas the model is practically indistinguishable from standard $\Lambda$CDM for $z\gtrsim 3.5$ (see~\cref{sec:appC} for a detailed discussion).

A companion multi-probe observational analysis of this reconstructed (sigmoid/quiescent) $\Lambda_{\rm s}(a)$ realisation given in~\cref{eqn:smoothLs},
including CMB, BAO, SN~Ia, and cosmic-shear data, has been presented in Ref.~\cite{Akarsu:2024eoo},
where $\Lambda_{\rm s}$VCDM was found to improve the overall fit and to alleviate the $H_0$ and $S_8$ tensions.

\section{Conclusions}

We have successfully integrated the $\Lambda_{\rm s}$CDM model~\cite{Akarsu:2019hmw, Akarsu:2021fol, Akarsu:2022typ, Akarsu:2023mfb} into a novel theoretical framework by endowing it with a specific Lagrangian, that of a type-II minimally modified gravity dubbed the VCDM model~\cite{DeFelice:2020eju, DeFelice:2020cpt}. Consequently, $\Lambda_{\rm s}$CDM has now acquired the status of a fully predictive cosmological model, applicable at any time and in any configuration, even beyond the cosmological context. In particular, we now have a fully predictive model for the entire evolution of the universe, including during the AdS-to-dS transition epoch. This enhancement also removes any ambiguity in applying the $\Lambda_{\rm s}$CDM model when confronting it with observational data.

In the VCDM framework, we have demonstrated that, most generally, an auxiliary scalar field $\phi$ endowed with a linear potential induces an effective cosmological constant (CC), $\Lambda_{\rm eff}=\text{const}$, in the Friedmann equation. This finding has enabled us to realise an abrupt mirror AdS-to-dS transition, as proposed in the \textit{abrupt} $\Lambda_{\rm s}$CDM scenario~\cite{Akarsu:2021fol,Akarsu:2022typ,Akarsu:2023mfb}, by defining two types of piecewise linear potentials with two segments---namely, a continuous piecewise-linear form and a discontinuous, flat form. To eliminate the type II (sudden) singularity~\cite{Barrow:2004xh}, which arises at the instant of an abrupt transition~\cite{Paraskevas:2024ytz}, and to ensure stable transitions, we have smoothed these piecewise potentials by replacing the junctions with a blended sigmoid interpolant. Our work reveals that the \textit{smooth} $\Lambda_{\rm s}$CDM model realised within the VCDM theory, here referred to as the $\Lambda_{\rm s}$VCDM theory, facilitates two types of rapid, smooth mirror AdS-to-dS transitions of the effective CC ($\Lambda_{\rm s}: \Lambda_{\rm AdS}<0\!\to\!\Lambda_{\rm dS}>0$ with $\Lambda_{\rm dS}=|\Lambda_{\rm AdS}|$):
\begin{itemize}
\item \textit{Agitated mirror AdS-to-dS transition}, wherein the scalar field evolves in a potential that climbs smoothly and monotonically from a pre-transition flat AdS plateau to a post-transition flat dS plateau of equal magnitude, across a finite transition layer centred at $\phi=\phi_{\rm c}$ (the midpoint of the interpolant). During this interval, $\Lambda_{\rm s}$ undergoes a smooth mirror AdS-to-dS transition and develops a single central bump that, if the transition has completed by today, overshoots the late-time dS plateau. In our fiducial example (cf.~\cref{fig:VL_tanh}), the universe undergoes accelerated expansion ($\ddot a>0$) not only today (for $z\lesssim0.6$, as in standard $\Lambda$CDM), but also over a short additional interval within the transition, around $z\sim2$. Moreover, if the transition is sufficiently sharp (with width $\sim\eta^{-1}$)---i.e.\ if $\eta^{2}\Lambda_{\rm dS}>\sqrt{3}/2$, equivalently $\eta\gtrsim0.64\,H_0^{-1}$ for a flat fiducial cosmology with $\Omega_{\rm m0}\simeq0.3$---then the Hubble rate $H(z)$ can also develop a bump (a brief super--acceleration episode with $\dot H>0$, while keeping both the background and perturbations stable, with the theory embedded in VCDM) nearly coincident with that in $\Lambda_{\rm s}$, with an amplitude that grows as the transition is sharpened; this transient feature is nested within the overall transition and occurs on the AdS side shortly after the onset of the transition.
\item \textit{Quiescent mirror AdS-to-dS transition}, wherein the scalar field evolves on a smooth potential obtained by sigmoid--smoothing a continuous, piecewise--linear parent whose \emph{slope} changes across a finite transition layer centred at $\phi=\phi_{\rm c}$, while the potential itself remains continuous and exhibits no offset jump. Correspondingly, $\Lambda_{\rm s}$ undergoes a smooth mirror AdS--to--dS transition in a comparatively gentle manner with respect to the agitated case: $\Lambda_{\rm s}(a)$ can remain monotone---either strictly, or in a globally monotone envelope while developing shallow, localised edge--shoulders near the entrance and exit of the transition window. In contrast to the \emph{agitated} construction---for which a transient bump (overshooting the late--time dS plateau) in $\Lambda_{\rm s}$ is generic once the transition has completed by today---a central bump in $\Lambda_{\rm s}$ is not automatic here and would require a sufficiently sharp and/or tuned profile. Dynamically, besides the usual late--time accelerated expansion ($\ddot a>0$ for $z\lesssim0.6$, as in standard $\Lambda$CDM), the universe can also exhibit an additional accelerated--expansion interval associated with the transition (e.g.\ around $z\sim1.6$ in \cref{fig:VL-sigmoid}), although this extra phase need not occur for sufficiently slow (broad) transitions. As in the agitated case, $H(a)$ need not remain monotone across the transition: a genuine super--acceleration sub--phase ($\dot H>0$), and hence a bump in $H(z)$, can occur; for the logistic quiescent blend this occurs \emph{iff} $\Delta\alpha\,\eta>2/3$. Once this threshold is crossed and a bump is present, its prominence is set primarily by the slope--mismatch parameter $\Delta\alpha$, and thus does not grow parametrically with $\eta$, unlike in the agitated transition. Finally, in the scale--factor--driven reconstructed quiescent family of \cref{fig:VL-recon}, one imposes a smooth signum--like $\Lambda_{\rm s}(a)$ via a hyperbolic--tangent profile (yielding a strictly monotone AdS--to--dS transition with no entrance/exit shoulders and no central bump) and reconstructs the corresponding $V(\phi)$~\footnote{During the revision of this work, a companion multi-probe observational analysis of the reconstructed
(sigmoid/quiescent) $\Lambda_{\rm s}(a)$ realisation discussed here was performed; see Ref.~\cite{Akarsu:2024eoo}.}. In this family, the appearance of an extra accelerated--expansion interval during the transition and the onset of super--acceleration are governed by separate steepness thresholds in $\tilde\eta$: while all cases exhibit the standard late--time acceleration ($z\lesssim0.6$), the slowest transition does not develop an additional accelerated phase around $z\sim z_\dagger$, and only the sharpest reconstructed example develops a nested super--acceleration phase (and hence a bump in $H$).
\end{itemize}
Although the \emph{quiescent} and \emph{agitated} constructions connect the same asymptotic effective--cosmological--constant plateaus---realising, in both cases, a mirror AdS--to--dS evolution of $\Lambda_{\rm s}$---their net effect away from the transition can be understood in the familiar $\Lambda_{\rm s}$CDM spirit: the pre-transition expansion rate is reduced while the late-time Hubble rate is increased relative to $\Lambda$CDM, and the parameters can be tuned so that the comoving distance to last scattering $D_M(z_\ast)$ remains essentially unchanged (see \cref{sec:appC}). That said, once the transition is given a finite width, the intermediate dynamics becomes more structured and potentially non-trivial: the evolution may include an additional accelerated-expansion ($\ddot a>0$) interval during the transition and, if the background enters a region where $V_{,\phi\phi}>2/3$, a genuine super--acceleration sub-phase ($\dot H>0$) accompanied by a bump in $H(z)$. In such cases, the expansion rate can temporarily exceed its $\Lambda$CDM counterpart over a limited redshift range around the transition, implying that the mechanism by which the scenario impacts cosmological tensions (most notably the $H_0$ tension) can be more subtle than in the abrupt-limit intuition and warrants dedicated investigation. These distinct transient expansion histories are therefore expected to leave different imprints in both the background and perturbation sectors---with the latter likely to be especially sensitive because the VCDM linearised equations deviate most strongly from their GR counterparts during the transition---and may thus enable observational probes of the transition epoch in the smooth $\Lambda_{\rm s}$VCDM realisations considered here.

It is worth noting that embedding the $\Lambda_{\rm s}$CDM scenario into the VCDM framework can have concrete and potentially distinctive consequences for structure formation. In particular, while the linear perturbation system is, in form, identical to that of $\Lambda$CDM, the scalar sector differs through the modified relation~\eqref{eqn:dotphi}, which depends explicitly on $\dot H$; moreover, $\dot H$ departs from its GR expression only when $V_{,\phi\phi}\neq 0$, i.e.\ predominantly during the transition epoch. For a sufficiently rapid yet smooth transition, one typically finds a temporary enhancement of the background quantity $|\dot H|/(N H^{2})$ during this finite window, accompanied by a non-negligible (and non-trivial) departure of the Hubble rate $H(z)$ from its $\Lambda$CDM counterpart. The resulting signatures---both in background observables and in growth-related probes---can therefore be confronted directly with data. At the same time, the Hubble function in this scenario can be arranged to coincide very closely with its $\Lambda$CDM counterpart both before and after the transition (as demonstrated in the abrupt $\Lambda_{\rm s}$CDM limit in Refs.~\cite{Akarsu:2021fol,Akarsu:2022typ,Akarsu:2023mfb}). After the transition, the background is effectively $\Lambda$CDM with a shifted late-time cosmological constant, while at sufficiently high redshift (e.g.\ $z\gtrsim 3.5$) dark energy is subdominant and the expansion becomes observationally very close to $\Lambda$CDM. This underlies the model’s ability to remain compatible with both high- and low-redshift distance data while addressing major cosmological tensions, most notably the $H_0$ tension (see~\cref{sec:appC}). It has also been shown that in the simplest implementation of the scenario---namely, the abrupt $\Lambda_{\rm s}$CDM background with GR perturbations---the modified expansion history can alleviate the $S_8$ tension by altering the $H(z)$ dependence in the perturbation equations~\cite{Akarsu:2022typ}, can at the same time ease the recently identified growth-index ($\gamma$) tension (a more than $4\sigma$ upward shift in $\gamma$ relative to the GR benchmark value $\sim0.55$~\cite{Linder:2005in,Wang:1998gt,Nguyen:2023fip,Specogna:2023nkq})~\cite{Akarsu:2025ijk,Escamilla:2025imi}, and can imprint distinctive features in the nonlinear matter power spectrum~\cite{Akarsu:2025nns}. In the $\Lambda_{\rm s}$VCDM setup developed here, the perturbation system is defined self-consistently across the transition (unlike earlier $\Lambda_{\rm s}$CDM studies where the transition itself was not explicitly modelled), allowing its impact on observables to be predicted. The modified relation~\eqref{eqn:dotphi} is necessarily different from its GR/$\Lambda$CDM counterpart, particularly during the transition epoch (cf.~\cref{eqn:dotphi}). As a result, different transition types and sharpness parameters are expected to yield quantitatively different predictions for $S_8$, $\gamma$, and related growth observables. A dedicated multi-probe analysis, combining all relevant datasets, will therefore be essential to rigorously test the scenario and to identify the subfamily of potentials/transitions that can simultaneously address major cosmological tensions such as  the $H_0$, $S_8$, and $\gamma$ tensions.

Thus, the integration of the $\Lambda_{\rm s}$CDM scenario into the VCDM framework elevates it to a fully predictive model, we referred to as the $\Lambda_{\rm s}$VCDM model, capable of describing the evolution of the Universe, including the late-time AdS-to-dS transition epoch. This advancement not only broadens the applicability of the $\Lambda_{\rm s}$CDM paradigm, but also has the potential to deepen our understanding of cosmological phenomena by providing a complete, self-consistent description of both the background and perturbation dynamics at all redshifts, thereby paving the way to reassess and potentially resolve the observational tensions plaguing the $\Lambda$CDM model. We believe further work is warranted to assess whether $\Lambda_{\rm s}$CDM can emerge as a credible extension of the concordance model, or at least as a useful guide for exploring its potential revisions.

\begin{acknowledgments}
\"{O}.A.\ acknowledges the support of the Turkish Academy of Sciences in the scheme of the Outstanding Young Scientist Award (T\"{U}BA-GEB\.{I}P). \"{O}.A.\ is supported in part by TUBITAK grant 122F124. The work of A.D.F. was supported by the Japan Society for the Promotion of Science Grants-in-Aid for Scientific Research No.~20K03969 and by grant PID2020-118159GB-C41 funded by MCIN/AEI/10.13039/501100011033. E.D.V. acknowledges support from the Royal Society through a Royal Society Dorothy Hodgkin Research Fellowship. S.K. gratefully acknowledges the support of Startup Research Grant from Plaksha University  (File No. OOR/PU-SRG/2023-24/08), and Core Research Grant from Science and Engineering Research Board (SERB), Govt. of India (File No.~CRG/2021/004658). R.C.N. thanks the financial support from the Conselho Nacional de Desenvolvimento Cient\'{i}fico e Tecnologico (CNPq, National Council for Scientific and Technological Development) under project No.\ 304306/2022-3, and the Funda\c{c}\~{a}o de Amparo \`{a} pesquisa do Estado do RS (FAPERGS, Research Support Foundation of the State of RS) for partial financial support under project No.\ 23/2551-0000848-3. E.\"{O}.~acknowledges the support by The Scientific and Technological Research Council of Turkey (T\"{U}B\.{I}TAK) in the scheme of 2214/A National PhD Scholarship Program. J.A.V.\ acknowledges the support provided by FOSEC SEP-CONACYT Investigaci\'on B\'asica A1-S-21925, UNAM-DGAPA-PAPIIT IN117723 and FORDECYT-PRONACES-CONACYT/304001/2019. A.Y.\ is supported by a Junior Research Fellowship (CSIR/UGC Ref.\ No.\ 201610145543) from the University Grants Commission, Govt. of India. This article/publication is based upon work from COST Action CA21136 -- ``Addressing observational tensions in cosmology with systematics and fundamental physics (CosmoVerse)'', supported by COST (European Cooperation in Science and Technology).
\end{acknowledgments}
\section*{Data Availability}
No new data were generated or analysed in support of this research.
\appendix

\section{Type-II minimally modified gravity and producing shifting effective cosmological constant}
\label{sec:appA}
In the type-II minimally modified gravity introduced in~\cite{DeFelice:2020eju} (later dubbed VCDM~\cite{DeFelice:2020cpt}), the standard cosmological constant, $\Lambda$, is promoted to a potential $V(\phi)$ of a non-dynamical auxiliary field, $\phi$, without introducing extra physical degrees of freedom. The action of this theory is given by
\begin{align}
    S&=S_{\rm m}+\!\Mpl^2\int d^4x\, N\sqrt{\gamma}\left[\frac{1}{2}\,\bigl(R+K_{ij}K^{ij}-K^2\bigr)\right.\nonumber\\
    &-\left.V(\phi)+\frac{\lambda_2}N  \gamma^{ij} D_i D_j \phi
    -\frac{3\lambda^2}{4}-\lambda(K+\phi)\right]\!,
\end{align}
where $S_{\rm m}$ is the sum of standard matter Lagrangians, $N$ is the lapse, and $K_{ij}$ is the extrinsic curvature (with $K=\gamma^{ij}K_{ij}$ as its trace) relative to the 3D metric $\gamma_{ij}$ (endowed with inverse $\gamma^{ij}$, determinant $\gamma$, and a covariant derivative $D_i$). Instead $\lambda$, $\lambda_2$, and $\phi$ are auxiliary fields. This modified gravity theory breaks four-dimensional diffeomorphism invariance but retains three-dimensional spatial diffeomorphism invariance and time-reparametrization invariance.~\footnote{As shown in Ref.~\cite{DeFelice:2022uxv}, by using the St\"uckelberg trick, it is possible to rewrite the theory in a fully covariant form. A time-like St\"uckelberg field, i.e., a chronon field, is present and acquires a time-like VEV on any background of the theory. On each of these backgrounds, we quantize all the fields following the  quantum field theory (QFT) procedures commonly employed on curved spacetimes~\cite{Mukhanov_Winitzki_2007}. This approach also applies to other minimal theories, such as the Cuscuton~\cite{Afshordi:2006ad}.} Consequently, on a homogeneous and isotropic background, it can modify the Hubble expansion rate while maintaining only two gravitational degrees of freedom, as in GR. In general, this allows for a spectrum of possibilities typically much broader than in scalar-tensor theories. For the latter, generally, the extra scalar degree of freedom imposes strong constraints, both locally (viz., at solar system scales) and at cosmological scales. To evade the local constraints, the scalar must be either very massive or shielded by non-trivial dynamical mechanisms, such as chameleon or Vainshtein effects. Additionally, constraining the cosmological background dynamics is necessary to avoid ghost and gradient instabilities.

The equations of motion for VCDM on a homogeneous and isotropic background can be expressed as~\cite{DeFelice:2020eju}:
\begin{equation}
\begin{aligned}
V = \frac{1}{3}\phi^2 -\frac{\rho}{\Mpl^2}\,,\quad\quad
\frac{{\rm d}\phi}{{\rm d}\mathcal{N}} = \frac{3}{2}\frac{\rho+P}{\Mpl^2H}\,,\\
\frac{{\rm d}\rho_I}{{\rm d}\mathcal{N}} +3(\rho_I +
P_I)=0\,,     
\end{aligned}
\label{eqn:backgroundEOM}
\end{equation}
where $\mathcal{N}=\ln(a/a_0)$, with $a_0=1$ being the present-day value of the scale factor, and $H=\dot{a}/(aN)$ is the Hubble expansion rate, with a dot denoting differentiation w.r.t. the time variable $t$, which is specified only after fixing $N$. For instance, when fixing $N=a$, the variable $t$ corresponds to the conformal time. It is important to note, as follows from~\cref{eqn:backgroundEOM}, that $\phi$ evolves monotonically with the scale factor: for $H>0$ (an expanding universe) and $\rho+P>0$ (matter content satisfying the null energy condition), one has ${\rm d}\phi/{\rm d}a>0$, so $\phi$ necessarily increases with cosmic time, i.e.\ as the universe expands. To derive the above equation, we have assumed a cosmological expansion history for which $H\neq0$, a condition expected to hold at least after inflation.~\footnote{Choosing a different time variable is sufficient to describe certain phenomena, such as a bouncing solution~\cite{Ganz:2022zgs}.} Additionally, we have defined $\rho=\sum_I\rho_I$ and $P=\sum_I P_I$, where the sum includes all standard matter species, including the dark matter component, each satisfying the usual continuity equation.

Provided that $\rho + P \neq 0$, the following equation can be derived from the set of equations presented in~\cref{eqn:backgroundEOM}:~\footnote{The theory is also endowed with a shadowy mode, another auxiliary field, whose profile on this background is given by $\lambda=-\frac23\,\phi-2H$.}
\begin{equation}
    \phi = \frac{3}{2}V_{,\phi} - 3H,\label{eq:phi_H}
\end{equation}
where $V_{,\phi}\equiv \dv{V}{\phi}$. This equation can be used to specify the initial condition for $\phi$ today: in our model, where the AdS-to-dS transition occurs in our past, $V(z=0)$ can be approximated by a linear function of $\phi$, namely $V \approx \alpha\phi + \beta$, thereby implying the initial condition $\phi(z=0)\approx\frac32\,\alpha-3H_0$, for a chosen value of $\alpha$. By combining the equations of motion, we express the Friedmann equation as follows;
\begin{equation}
3\Mpl^{2}H^{2}=\rho+\rho_{\phi}\,, \label{eqn:Friedmann-eq}
\end{equation}
where
\begin{equation}
\rho_{\phi}\equiv\Mpl^{2}(V-\phi V_{,\phi})+\frac{3}{4}\,\Mpl^{2}\,V_{,\phi}^{2}\,.
\label{eq:rho_phi}
\end{equation}
Taking derivatives of the expressions for $\phi$ and $H$ reveals that
\begin{equation}
\frac{\dot{H}}{N}=\frac{ ( \rho+P )\,( 3\,V_{{,\phi\phi}}-2 ) }{4\Mpl^2}\,,\label{eq:dot_H}
\end{equation}
which deviates from the corresponding equation in GR if and only if $V_{,\phi\phi} \ne 0$. If $V$ is a linear function of $\phi$, the standard $\Lambda$CDM dynamics are obtained. Examining~\cref{eq:dot_H}, we find that it can be expressed in terms of an effective pressure defined by
\begin{equation}
  P_\phi=-\frac32\,(\rho+P)\,V_{,\phi\phi}-\rho_\phi\,,\label{eq:P_phi}
\end{equation}
which makes it evident that $P_\phi\rightarrow-\rho_\phi$ (mimicking a CC, $\rho_\phi\to\rho_\Lambda=\rm const$) as $V_{,\phi\phi}\to0$ (approaching a linear function). It should be noted that by using the second equation in~\cref{eqn:backgroundEOM}, along with~\cref{eq:phi_H,eq:rho_phi,eq:P_phi}, we see that the auxiliary scalar field, i.e., the $\phi$ component, satisfies local energy-momentum conservation, viz., the continuity equation $\dot{\rho}_\phi/\mathcal{N}+3H(\rho_\phi+P_\phi)=0$. Importantly, since no physical particle, i.e., an additional degree of freedom, is associated with $\phi$, the apparent violation of the weak energy condition (WEC) indicated by $\dot{\rho}_\phi>0$ does not lead to classical or quantum instabilities.~\footnote{It should be noted that it is a key feature of our model that $\rho_\phi$ increases as the universe expands (as $z$ decreases), at least for a period of time, but this does not necessarily imply that the scalar field $\phi$ climbs up the potential $V(\phi)$ in VCDM theory. As can be seen, for instance, in~\cref{fig:VL-step,fig:VL-sigmoid,fig:VL-recon}, $\Lambda_{\rm s}=\rho_\phi/\Mpl^2$ increases as $\phi$ rolls down the potential $V(\phi)$.} This is because the $\phi$ field, unlike conventional matter fields, does not represent a physical substance that could contribute to such instabilities. As discussed in Refs.~\cite{DeFelice:2020eju, DeFelice:2020cpt}, this unique characteristic of the $\phi$ field in VCDM theory ensures the stability of the cosmological model, even in scenarios where the WEC condition might appear violated. Indeed, for scalar field to induce an effective cosmological constant, $\Lambda_{\rm eff}=\rm const$, in the Friedmann equation, solving~\cref{eq:rho_phi} reveals that:~\footnote{The solution $V(\phi)=\frac{\phi^2}{3}+\beta$, $\beta$ being a constant, should be excluded as~\cref{eq:phi_H} results in $H=0$, leading to a non-dynamical evolution of the Universe.}
\begin{equation}
    \rho_\phi/\Mpl^2=\Lambda_{\rm eff}\,\, \Rightarrow\,\, V(\phi)=\alpha\phi-\frac{3}{4}\alpha^2+\Lambda_{\rm eff}.
\end{equation}
This result implies that an abrupt shift in $\Lambda_{\rm eff}$ at a critical value of $\phi=\phi_{\rm c}$ from one value to another is not necessarily caused by a potential that has undergone an abrupt shift. For instance, it is always possible to define a continuous piecewise linear potential with two pieces, and as the scalar field rolls through it, it causes a shift in $\Lambda_{\rm eff}$ due to the sudden change in the slope at $\phi=\phi_{\rm c}$.

It is precisely these features of the VCDM~\cite{DeFelice:2020eju, DeFelice:2020cpt} that enable us to realize the $\Lambda_{\rm s}$CDM model~\cite{Akarsu:2019hmw, Akarsu:2021fol, Akarsu:2022typ, Akarsu:2023mfb} within this framework, offering two main types of scenarios. In particular, an abrupt mirror AdS-to-dS transition, (namely, an abrupt shift in $\Lambda_{\rm eff}$ from $\Lambda_{\rm AdS} < 0$ to $\Lambda_{\rm dS} = -\Lambda_{\rm AdS} > 0$) at $\phi=\phi_{\rm c}$, is not necessarily caused by a potential that has undergone an abrupt shift from $V_{\rm b}(\phi) = -\Lambda_{\rm dS} < 0$ to $V_{\rm a}(\phi) = \Lambda_{\rm AdS} > 0$. Instead, in the general scenario, it can be caused by a potential transitioning between two separate linear regimes, namely, from $V_{\rm b}(\phi) = \alpha_{\rm b}\phi - \frac{3}{4}\alpha_{\rm b}^2 - \Lambda_{\rm dS}$ for $\phi<\phi_{\rm c}$ to $V_{\rm a}(\phi) = \alpha_{\rm a}\phi - \frac{3}{4}\alpha_{\rm a}^2 + \Lambda_{\rm dS}$ for $\phi\geq\phi_{\rm c}$, where $\phi_{\rm c}<\phi(z=0)$. That is, an abrupt sign-switch can occur due to either \textbf{(i)} an abrupt jump in the value of the potential, \textbf{(ii)} a sudden change in the slope of the potential, or \textbf{(iii)} a combination of the first two. In neither of these cases is the potential smooth; if $V_{\rm a}(\phi=\phi_{\rm c})=V_{\rm b}(\phi=\phi_{\rm c})$, it is continuous, but not in its derivatives. If $V_{\rm a}(\phi=\phi_{\rm c})\neq V_{\rm b}(\phi=\phi_{\rm c})$, even the potential itself is discontinuous. Note, on the other hand, that in any case, the Hubble function, $H$, would be discontinuous at $\phi=\phi_{\rm c}$, leading to a type II (sudden) singularity in the past~\cite{Barrow:2004xh,Paraskevas:2024ytz}.

To circumvent this issue, one can straightforwardly devise a smooth potential by implementing a blended function [see, e.g.,~\cref{eq:smooth_potential}] that ensures a smooth transition between two linear regimes by incorporating a sigmoid function, specifically, a potential that approximates $V(\phi)\approx V_{\rm b}(\phi)$ for $\phi<\phi_{\rm c}-\Delta\phi/2$ and transitions to $V(\phi)\approx V_{\rm a}(\phi)$ for $\phi>\phi_{\rm c}+\Delta\phi/2$. Here, $\Delta\phi$ denotes the breadth of the transition epoch centred around $\phi_{\rm c}$---in the case of~\cref{eq:smooth_potential} or similar potentials for which $\rho_\phi$ evolves at all times, the transition epoch can be defined as the period between the first violation of the condition $\abs{\Lambda_{\rm dS}-\Lambda_{\rm s}(a)}/\Lambda_{\rm dS}<\varepsilon$ until the end of the last violation for a chosen small value of $\varepsilon$. The post- and pre-transition eras correspond to when the two branches of the sigmoid function resemble almost straight lines. In these eras, the effective cosmological constant, $\Lambda_{\rm eff}$, is indistinguishable from a constant phenomenologically. Currently, as we are in the post-transition era, it is reasonable to assume that $\Lambda_{\rm eff}(z=0) = \Lambda_{\rm s0} \approx \Lambda_{\rm dS}$, akin to the abrupt $\Lambda_{\rm s}$CDM scenario~\cite{Akarsu:2019hmw, Akarsu:2021fol, Akarsu:2022typ, Akarsu:2023mfb}. For example, in the logistic function provided in~\cref{eqn:logisticF}, $S(\phi)$ approaches 0 or 1 at the 99 percent level for $\phi = \phi_{\rm c} \pm 2\eta^{-1}$. Using this approximation, the width of the transition region can be identified as $\Delta\phi\approx 4\eta^{-1}$, or equivalently in dimensionless form as $\Delta\phi/|\phi_{\rm c}| \approx 4\zeta^{-1}$.

Now, for instance, incorporating aforementioned linear potentials into the blended function given in~\cref{eq:smooth_potential}, the resultant potential can be expressed as:
\begin{align}
    V(\phi) &= 
        [\alpha_{\rm b} (\phi-\phi_{\rm c})-\frac34\,\alpha_{\rm b}^2+\alpha_{\rm b}\phi_{\rm c}-\Lambda_{\rm dS}]\,[1-S(\phi)]\nonumber\\
        &+[\alpha_{\rm a} (\phi-\phi_{\rm c})-\frac34\,\alpha_{\rm a}^2+\alpha_{\rm a}\phi_{\rm c}+\Lambda_{\rm dS}]\,S(\phi),
    \label{eq:gen_sigmoid_pot}
\end{align}
where $S(\phi)$ represents a sigmoid function, e.g., the logistic function given in~\cref{eqn:logisticF}. This formulation leads us to identify two possible types of smooth mirror AdS-to-dS transition, based on whether the piecewise linear potential with two pieces to be smoothed out is continuous or discontinuous, namely, whether the original function satisfies $V_b(\phi=\phi_{\rm c}) \neq V_a(\phi=\phi_{\rm c})$ or not:
\begin{enumerate}[nosep,wide,label=\roman*.]
\item
\textit{Agitated mirror AdS-to-dS transition}: In this type, the piecewise linear potential with two pieces to be smoothed out is discontinuous since $V_b(\phi=\phi_{\rm c}) \neq V_a(\phi=\phi_{\rm c})$. It is necessary to smooth out not only the possible sudden change in slope at $\phi=\phi_{\rm c}$, but also the jump in the potential at this point. Consequently, the smoothing process will reflect the discontinuity of the original potential. Specifically, for a sufficiently fast transition, $|V_{,\phi}|$ and $|V_{,\phi\phi}|$ will tend to large values within the transition region, resulting in significant increases in both $H^2/H_0^2$ and $|\dot{H}|/(NH^2)$ during the transition.
\item
\textit{Quiescent mirror AdS-to-dS transition}: In this type, the piecewise linear potential with two pieces to be smoothed out is continuous; $V_b(\phi=\phi_{\rm c}) = V_a(\phi=\phi_{\rm c})$. Unlike the agitated transition, there is no jump in the original potential to address; it is solely the sudden change in the slope of the potential at $\phi=\phi_{\rm c}$ to be smoothed out, and the change in the slope need not be large. Consequently, $|V_{,\phi}|$ generically remains controlled and does not necessarily increase substantially during the transition. However, $|V_{,\phi\phi}|$ might be large, contributing to significant increases in $|\dot{H}|/(NH^2)$ during the transition. For the quiescent mirror AdS-to-dS transition, the potential~\eqref{eq:gen_sigmoid_pot} requires the following additional relations:
\begin{equation}
    \phi_{\rm c}=\frac{3\alpha_{\rm b}^2-3\alpha_{\rm a}^2+8\Lambda_{\rm s0}}{4(\alpha_{\rm b}-\alpha_{\rm a})}<\phi_0=\frac{3}{2}\alpha_{\rm a}-3H_0\,,
\end{equation}
where $\phi_0=\phi(z=0)$. Assuming also that $H_0^2>\Lambda_{\rm s0}/3>0$, we identify two possible sets of parameter solutions. The first is
\begin{equation}
    \mathcal{A}_1 = \left\{H_{0}\le \frac{\alpha_{{\rm a}}}{4}-\frac{\alpha_{{\rm b}}}{4}, \alpha_{{\rm b}}<\alpha_{{\rm a}}\right\} .
\end{equation}
And, the second set of solutions is given by
\begin{align}   
\mathcal{A}_2 &=\left\{\frac{\alpha_{{\rm a}}}{4}-\frac{\alpha_{{\rm b}}}{4}<H_{0}, \alpha_{{\rm b}}<\alpha_{{\rm a}},\right.\nonumber\\
&\left. \quad\frac{3}{2} H_{0} (\alpha_{{\rm a}}-\alpha_{{\rm b}})-\frac{3}{8}  (\alpha_{{\rm a}}-\alpha_{{\rm b}})^2<\Lambda_{\rm s0}
\right\}.
\end{align}
In both cases, $\alpha_{{\rm b}}<0$ for $\alpha_{{\rm a}}=0$.
\end{enumerate}

\section{Behaviour of $\Lambda_{\mathrm{s}}(a)$ during the transition}
\label{sec:appFeature}

We now elaborate on the behaviour of the effective cosmological constant, $\Lambda_{\rm s}$, during the transition epoch for the \emph{Quiescent} and \emph{Agitated} mirror AdS--to-dS transition scenarios. Using~\cref{eq:phi_H,eqn:Friedmann-eq,eq:rho_phi}, along the homogeneous background one finds
\begin{equation}
\Lambda_{{\rm s},\phi}
=\frac{1}{M_{\rm Pl}^2}\,\partial_\phi\rho_\phi
=\Big(\tfrac{3}{2}\,V_{,\phi}-\phi\Big)V_{,\phi\phi}
=3H\,V_{,\phi\phi}\;,
\end{equation}
which implies
\begin{equation}
\frac{{\rm d}\Lambda_{\rm s}}{{\rm d}a}
=3H\,V_{,\phi\phi}\,\frac{{\rm d}\phi}{{\rm d}a}\,.
\label{eq:Lambda_s_chainrule}
\end{equation}
Hence, in an expanding background (${H>0}$) where the scalar field evolves monotonically with the scale factor (in our realizations ${\rm d}\phi/{\rm d}a>0$ as already explained after~\cref{eqn:backgroundEOM}), the \emph{sign} of ${\rm d}\Lambda_{\rm s}/{\rm d}a$ is controlled solely by the curvature $V_{,\phi\phi}$ sampled along the trajectory $\phi(a)$. In particular:
(i) $\Lambda_{\rm s}(a)$ is \emph{strictly monotone} across the transition provided $V_{,\phi\phi}(\phi(a))$ does not change sign on the transition interval (and does not vanish identically on any subinterval); 
(ii) each sign flip of $V_{,\phi\phi}$ along $\phi(a)$ produces one localised extremum of $\Lambda_{\rm s}(a)$.
In the idealized piecewise--linear limit, $V_{,\phi\phi}=0$ away from the transition-layer centre $\phi_{\rm c}$ and the slope discontinuity at $\phi_{\rm c}$ yields a step in $\Lambda_{\rm s}$ (an abrupt jump at the transition). Any finite--width regularization that remains everywhere \emph{strictly convex} or \emph{strictly concave} over the transition window keeps $\Lambda_{\rm s}(a)$ strictly monotone. By contrast, a smooth $V(\phi)$ that interpolates between the two linear regimes via an inflection-bearing sigmoid generically induces small, localised departures from strict monotonicity near the entrance and exit of the transition window (or, for certain constructions, centrally), as detailed below.

Before focusing on particular transition types, it is useful to make the general structure explicit. To this end, we write the interpolating potential in the generic form
\begin{equation}
    V(\phi)=B(\phi)\big[1-S(\phi)\big]\,+\,A(\phi)\,S(\phi),
\end{equation}
where $B(\phi)$ and $A(\phi)$ are \emph{affine} functions (linear in $\phi$ with constant offsets) with slopes $\alpha_{\rm b}$ and $\alpha_{\rm a}$, and $S(\phi)$ is a smooth interpolant that switches monotonically from $0$ to $1$ across the transition region. We denote by $\phi_{\rm c}$ the transition-layer centre, defined by $S(\phi_{\rm c})=\tfrac12$ (equivalently, for single--inflection sigmoids, the unique zero of $S''$). We require $S\in C^2$ (at least), i.e.\ twice continuously differentiable, so that both its slope $S'$ and curvature $S''$ are well defined; this includes the standard sigmoid choices used in practice (e.g.\ logistic, error function). With this setup, one finds
\begin{align}
V_{,\phi}(\phi)&=\alpha_{\rm b}\;+\;\Delta\alpha\,S(\phi)\;+\;\Delta(\phi)\,S'(\phi), \label{eq:Vp_general}\\
V_{,\phi\phi}(\phi)&=2\,\Delta\alpha\,S'(\phi)\;+\;\Delta(\phi)\,S''(\phi),
\label{eq:Vpp_generalS}
\end{align}
where $\Delta\alpha\equiv \alpha_{\rm a}-\alpha_{\rm b}$ is the slope contrast between the two linear branches, and $\Delta(\phi)\equiv A(\phi)-B(\phi)$ is an affine offset, which (for affine parents) satisfies $\Delta'(\phi)=\Delta\alpha$. Whether the underlying parent (piecewise--linear) construction is continuous at the centre is encoded by
$\Delta(\phi_{\rm c})=A(\phi_{\rm c})-B(\phi_{\rm c})$.
In the \emph{quiescent} subclass we impose $\Delta(\phi_{\rm c})=0$ (continuity of the parent potential, only the slope changes), whereas in the \emph{agitated} subclass we deliberately take $\Delta(\phi_{\rm c})\neq0$ (a jump at $\phi_{\rm c}$), which remains after smoothing as the driver of a single central feature in $\Lambda_{\rm s}(a)$ (cf.~\cref{sec:appA}). Accordingly, two regimes follow naturally:
\begin{enumerate}[nosep,wide,label=\roman*.]
\item
\textit{Strictly monotone case.} If $V_{,\phi\phi}$ keeps a definite sign along the portion of $\phi$ traversed by the background, then $\Lambda_{\rm s}(a)$ is strictly monotone. A convenient sufficient condition is $\big|\,\Delta(\phi)\,S''(\phi)\,\big| \;<\; 2\,|\Delta\alpha|\,S'(\phi)$ throughout the transition interval (where $S'(\phi)>0$ for standard sigmoids), in which case $V_{,\phi\phi}$ retains the sign of $\Delta\alpha$. This is naturally satisfied for sufficiently broad, gentle blends (i.e.\ small $|\phi_{\rm c} S''|/S'$ across the transition; for the logistic interpolant this is governed by $\zeta\equiv\eta|\phi_{\rm c}|$ at fixed $\phi_{\rm c}$ in our convention). \emph{Example (reconstructed).} In the reconstruction setup (cf.\ \cref{fig:VL-recon}), the background history is prescribed and $V(\phi)$ is inferred along the realized trajectory. Evaluated on $\phi(a)$, the combination in~\cref{eq:Vpp_generalS} keeps a definite sign in our solutions; by~\cref{eq:Lambda_s_chainrule}, ${\rm d}\Lambda_{\rm s}/{\rm d}a$ then has fixed sign (for $H>0$ and monotone $\phi$), so $\Lambda_{\rm s}(a)$ remains strictly monotone despite possibly large $|V_{,\phi\phi}|$.
\item
\textit{Localised non--monotonic features.} Whenever the curvature term $\Delta(\phi)\,S''$ overtakes the slope--contrast term $2\,\Delta\alpha\,S'$ somewhere along the trajectory, $V_{,\phi\phi}$ necessarily vanishes and changes sign, and $\Lambda_{\rm s}(a)$ develops localised extrema. For typical single--inflection sigmoids, $S'$ is an even, positive function peaked at the centre ($\phi_{\rm c}$), while $S''$ is odd and changes sign once. The slope--contrast contribution $2\,\Delta\alpha\,S'$ is therefore symmetric and largest near the centre, whereas the curvature contribution $\Delta(\phi)\,S''$ is antisymmetric, vanishing at the centre but growing in magnitude toward the edges. Their competition causes $V_{,\phi\phi}$ to cross zero twice---once on each side of the centre---so that $\Lambda_{\rm s}(a)$ develops two extrema near the entrance and exit of the transition window, i.e.\ the ``edge--shoulder'' structure characteristic of quiescent blends (cf.\ \cref{fig:VL-sigmoid}). The strength of any such feature reflects the balance between the slope--contrast piece $\propto \Delta\alpha\,S'$ and the curvature piece $\propto \Delta(\phi)\,S''$ in~\cref{eq:Vpp_generalS}, modulated by the background sweep factor $3H\,{\rm d}\phi/{\rm d}a$ in~\cref{eq:Lambda_s_chainrule}. More precisely, for $\Delta\alpha\neq0$ the zero--crossing condition for $V_{,\phi\phi}$ can be written in a dimensionless form by dividing~\cref{eq:Vpp_generalS} by $S'(\phi)>0$ and defining ${\mathcal{R}(\phi)\,\equiv\,[\Delta(\phi)/\Delta\alpha]\,[S''(\phi)/S'(\phi)]}$. Non--monotonic features arise iff $2+\mathcal{R}(\phi)$ changes sign somewhere along the trajectory (i.e., $\min_{\rm path}[\,2+\mathcal R(\phi)\,]<0$). Thus $\Delta(\phi)$ \emph{does} enter through the lever--arm $\Delta(\phi)/\Delta\alpha$, while the interpolant’s rapidity is captured by the dimensionless ratio $|\phi_{\rm c}S''|/S'$ (for the logistic interpolant, $S''/S'=-2\eta\tanh[\eta(\phi-\phi_{\rm c})]$, so this is governed by $\zeta$, as $|\phi_{\rm c}S''|/S'=2\zeta\,|\tanh[\eta(\phi-\phi_{\rm c})]|$). In the quiescent constructions with continuity at the centre, $A(\phi_{\rm c})=B(\phi_{\rm c})$, one has $\Delta(\phi)=\Delta\alpha\,(\phi-\phi_{\rm c})$, so ${\mathcal{R}(\phi)=(\phi-\phi_{\rm c})\,S''/S'}$, which for the logistic interpolant reaches $|\mathcal R|\simeq 4$ near the transition edges (e.g., $|\phi-\phi_{\rm c}|\approx 2/\eta$), making the edge shoulders generic whenever the trajectory spans the full window. These features are suppressed whenever $|\mathcal{R}|\ll 2$ across the sampled interval---for example, for gentle sigmoids (small $|\phi_{\rm c}S''|/S'$), small slope contrast $|\Delta\alpha|$, or a small offset--to--slope ratio $|\Delta(\phi)/\Delta\alpha|$; conversely, features are enhanced once $|\mathcal{R}|$ reaches $\mathcal{O}(1)$--few.
\end{enumerate}

\emph{Matched--slope special case ($\Delta\alpha=0$).} In addition to the two regimes above, when the branch slopes match the discussion simplifies. Because $B(\phi)$ and $A(\phi)$ are affine, two distinct outcomes occur. If $B(\phi)=A(\phi)$, then $V(\phi)=B(\phi)$ for all $\phi$ and $V_{,\phi\phi}\equiv 0$, so $\Lambda_{\rm s}$ is constant and no transition occurs. If instead $B(\phi)=m\phi+c_{\rm b}$ and $A(\phi)=m\phi+c_{\rm a}$ with $c_{\rm a}\neq c_{\rm b}$, then $\Delta(\phi)=\Delta V\equiv c_{\rm a}-c_{\rm b}$ is constant and $V(\phi)=m\phi+c_{\rm b}+\Delta V\,S(\phi)$ with $V_{,\phi\phi}=\Delta V\,S''(\phi)$. Note that continuity at the centre would require $\Delta V=\Delta(\phi_{\rm c})=0$, which collapses this case to the trivial no--transition outcome $V_{,\phi\phi}\equiv0$.
Thus the nontrivial matched--slope realization necessarily has $\Delta V\neq0$ and corresponds to the \emph{agitated} subclass. For a single--inflection sigmoid $S$, $S''$ is odd with a single zero at the centre, so $V_{,\phi\phi}$ changes sign once and $\Lambda_{\rm s}(a)$ develops a single central extremum (the \emph{agitated} pattern; cf.\ \cref{fig:VL_tanh}). As a canonical mirror AdS-to-dS example with flat plateaus ($m=0$), one may take $V(\phi)=-\Lambda_{\rm dS}+2\Lambda_{\rm dS}\,S(\phi)$, for which $\Lambda_{\rm s}(a)$ exhibits a single central extremum (cf.\ \cref{fig:VL_tanh}).

The general criteria above organize the three constructions used in the main text: (a) in \emph{agitated} cases ($\Delta\alpha=0$ with $\Delta(\phi_{\rm c})\neq0$ from distinct offsets), $V_{,\phi\phi}\propto S''$ changes sign once and $\Lambda_{\rm s}(a)$ generically exhibits a single, central bump; (b) in \emph{quiescent} blends ($\Delta\alpha\neq 0$ with $\Delta(\phi_{\rm c})=0$), the competition between $2\Delta\alpha\,S'$ and $\Delta(\phi)\,S''$ can yield two edge shoulders provided the trajectory spans the transition window; however, this is \emph{not} guaranteed---strict monotonicity obtains whenever the curvature term never overtakes the slope--contrast term; (c) the \emph{reconstructed} example explicitly illustrates this monotone outcome along the realized trajectory.

Localised non--monotonicity in $\Lambda_{\rm s}(a)$ (edge shoulders or a central bump) propagates into percent--level features in the late--time expansion rate $E(z)$ concentrated around the transition redshift $z_\dagger$. Current low--$z$ BAO and cosmic--chronometer data disfavour large, sharp departures, which naturally steer quiescent realizations toward broader, gentler blends (small $|\phi_{\rm c} S''|/S'$ or smaller $|\Delta\alpha|$) and agitated realizations toward modest curvature concentration. Conversely, strictly monotone cases---such as the reconstructed example---are less constrained by such localised signatures.

\section{Super--acceleration phase (bump) in the Hubble parameter}
\label{app:bump}

Building on \cref{sec:appFeature}, we investigate when the Hubble rate exhibits a temporary rise ($\dot H>0$) during the mirror AdS–to–dS transition. Using $H=\tfrac12 V_{,\phi}-\tfrac{\phi}{3}$ together with \cref{eq:Vp_general}, the background can be written, for any single–inflection, monotone interpolant $S(\phi)$,
\begin{equation}
H(\phi)=\frac{\alpha_{\rm b}}{2}\;+\;\frac{\Delta\alpha}{2}\,S(\phi)\;+\;\frac{1}{2}\,\Delta(\phi)\,S'(\phi)\;-\;\frac{\phi}{3},
\label{eq:H_general}
\end{equation}
where $\Delta\alpha$ and $\Delta(\phi)$ are as defined previously (for affine parents, $\Delta'(\phi)=\Delta\alpha$). Differentiating $H(\phi)$ gives $H_{,\phi}=\tfrac12 V_{,\phi\phi}-\tfrac13$, and using~\cref{eq:Vpp_generalS} one obtains
\begin{equation}
H_{,\phi}=\Delta\alpha\,S'(\phi)\;+\;\tfrac12\,\Delta(\phi)\,S''(\phi)\;-\;\tfrac13,
\,\,\,
\dot H=H_{,\phi}\,\dot\phi.
\label{eq:Hphi_general}
\end{equation}
Along our expanding, monotone backgrounds ($H>0$, $\dot\phi>0$), $\dot H>0$ is equivalent to $V_{,\phi\phi}>\tfrac{2}{3}$. Interior turning points of $H(\phi)$ satisfy $V_{,\phi\phi}(\phi)=\tfrac{2}{3}$. Thus, super–acceleration occurs precisely on field intervals where $V_{,\phi\phi}>\tfrac{2}{3}$; a local bump in $H$ exists \emph{iff} there is a finite interval $(\phi_-,\phi_+)$ with $V_{,\phi\phi}>\tfrac{2}{3}$ and $V_{,\phi\phi}(\phi_\pm)=\tfrac{2}{3}$, so that $H_{,\phi}>0$ (equivalently, $\dot H>0$) inside and changes sign at the endpoints.

We proceed with the logistic interpolant introduced in~\cref{eqn:logisticF} and, for convenience, define ${u\equiv\eta(\phi-\phi_{\rm c})}$, where $\phi_{\rm c}$ is the transition-layer centre (the midpoint of the interpolant, $S(\phi_{\rm c})=\tfrac12$), and ${\phi_{\rm c}=-\zeta/\eta}$ (the minus sign reflects that ${\zeta\equiv\eta|\phi_{\rm c}|>0}$ while ${\phi_{\rm c}<0}$ in our realizations). From~\cref{eqn:logisticF} it follows that ${S'(\phi)=\tfrac{\eta}{2}\sech^{2}u}$ and ${S''(\phi)=-\eta^{2}\sech^{2}u\,\tanh u}$. For reference, in terms of $u$, the operational transition window (in practice, from the onset to the end of the AdS–to–dS transition) corresponds to $u\in[-2,2]$, which covers the portion of the transition where $S'$ and $S''$ are appreciable ($S(-2)\simeq0.018$ to $S(2)\simeq0.982$); the centre ${\phi=\phi_{\rm c}}$ maps to $u=0$. (Note that the zero-crossing of $\Lambda_{\rm s}$ need not occur exactly at $u=0$.)\\

\noindent \textit{Quiescent smooth transition:} For the quiescent blend (continuous parent with ${\Delta(\phi_{\rm c})=0}$ so that ${\Delta(\phi)=\Delta\alpha(\phi-\phi_{\rm c})=\Delta\alpha\,u/\eta}$), inserting ${S=\tfrac12(1+\tanh u)}$ and ${S'=\tfrac{\eta}{2}\sech^{2}\!u}$ into \cref{eq:H_general} yields
\begin{equation}
H(\phi)=\frac{\Sigma\alpha}{4}
+\frac{\Delta\alpha}{4}\Big[u+\tanh u - u\tanh^{2}u\Big]
+\frac{\zeta-u}{3\eta},
\label{eq:Hq_quiescent_exact}
\end{equation}
with $\Sigma\alpha\equiv\alpha_{\rm a}+\alpha_{\rm b}$. Differentiating, or equivalently using \cref{eq:Hphi_general} with $S''=-\eta^{2}\sech^{2}\!u\,\tanh u$, one obtains across the transition
\begin{equation}
H_{,\phi}=\frac12\,\Delta\alpha\,\eta\,\sech^{2}\!u\,\Big(1-u\tanh u\Big)-\frac{1}{3}.
\label{eq:Hphi_quiescent}
\end{equation}
Evaluating \cref{eq:Hq_quiescent_exact} at $u=-2,0,+2$ gives
\begin{equation}
\begin{aligned}
H_{\rm on}&=\frac{\Sigma\alpha}{4}
+\frac{\Delta\alpha}{4}\Big[-2-\tanh 2 + 2\tanh^{2}2\Big]
+\frac{\zeta+2}{3\eta},\\[2pt]
H_{\rm ex}&=\frac{\Sigma\alpha}{4}+\frac{\zeta}{3\eta},\\[2pt]
H_{\rm end}&=\frac{\Sigma\alpha}{4}
+\frac{\Delta\alpha}{4}\Big[\,2+\tanh 2 - 2\tanh^{2}2\Big]
+\frac{\zeta-2}{3\eta}.
\end{aligned}
\label{eq:Hq_endpoints}
\end{equation}
Interior turning points of $H$ satisfy $H_{,\phi}=0$, i.e.
\begin{equation}
\Delta\alpha\,\eta\,\sech^{2}\!u\,\Big(1-u\tanh u\Big)=\frac{2}{3}.
\label{eq:root_quiescent}
\end{equation}
Since $\sech^{2}\!u\,(1-u\tanh u)\le 1$ with equality at $u=0$, interior solutions exist \emph{iff}
\begin{equation}
\Delta\alpha\,\eta>\frac{2}{3}.
\label{eq:threshold_quiescent_logistic}
\end{equation}
When \cref{eq:threshold_quiescent_logistic} holds there are two roots $u_-<0<u_+$; in the sharp regime $\Delta\alpha\,\eta\gg1$ they approach
$u_\pm\,\to\,\pm u_0$, where $u_0>0$ solves $u_0\tanh u_0=1$ ($u_0\simeq 1.19968$), so the AdS side of the transition gives a local minimum and the dS side a local maximum (as $\Delta\alpha>0$). Writing $\mathcal{S}(u)\equiv u+\tanh u-u\tanh^{2}u$, the exact interior heights follow from \cref{eq:Hq_quiescent_exact}:
\begin{equation}
\begin{aligned}
H_{\max}-H_{\rm ex}&=\frac{\Delta\alpha}{4}\,\mathcal{S}(u_+)-\frac{u_+}{3\eta},\\
H_{\rm ex}-H_{\min}&=\frac{\Delta\alpha}{4}\,\mathcal{S}(-u_-)+\frac{u_-}{3\eta},\\
H_{\max}-H_{\min}&=\frac{\Delta\alpha}{4}\Big[\mathcal{S}(u_+)-\mathcal{S}(u_-)\Big]-\frac{u_+-u_-}{3\eta}.
\end{aligned}
\label{eq:heights_quiescent_exact}
\end{equation}
In the sharp regime, using $\tanh u_0=1/u_0$ (so $\mathcal{S}(\pm u_0)=\pm u_0$) gives
\begin{equation}\label{eq:heights_quiescent_asym}
\begin{aligned}
H_{\max}-H_{\rm ex}\,\simeq\,u_0\!\left(\frac{\Delta\alpha}{4}-\frac{1}{3\eta}\right),\\
H_{\rm ex}-H_{\min}\,\simeq\,u_0\!\left(\frac{\Delta\alpha}{4}-\frac{1}{3\eta}\right),\\
H_{\max}-H_{\min}\,\simeq\,u_0\!\left(\frac{\Delta\alpha}{2}-\frac{2}{3\eta}\right),
\end{aligned}
\end{equation}
i.e.\ essentially linear growth with the slope mismatch ($\Delta\alpha$), with $1/\eta$ subtractions from the $-\phi/3$ drift. Endpoint comparisons are likewise
\begin{equation}
\begin{aligned}
H_{\max}-H_{\rm on}&=\frac{\Delta\alpha}{4}\Big[\mathcal{S}(u_+)+2+\tanh 2-2\tanh^{2}2\Big]\\
&\quad+\frac{-u_+-2}{3\eta},\\
H_{\max}-H_{\rm end}&=\frac{\Delta\alpha}{4}\Big[\mathcal{S}(u_+)-2-\tanh 2+2\tanh^{2}2\Big]\\
&\quad+\frac{-u_++2}{3\eta},
\end{aligned}
\label{eq:overshoot_quiescent}
\end{equation}
with analogous expressions for undershoots using $u_-$. If $\Delta\alpha\,\eta\le 2/3$, \cref{eq:root_quiescent} admits no interior solution; then $H_{,\phi}\le0$ across the window and $H(\phi)$ is strictly monotone (no bump). Regardless of \cref{eq:threshold_quiescent_logistic}, the factor $1-u\tanh u$ in \cref{eq:Hphi_quiescent} flips sign at $|u|=u_0$, producing shallow edge shoulders in $\Lambda_{\rm s}(\phi)$ (and only inflection–like features in $H$), as discussed in \cref{sec:appFeature}.\\

\noindent {\textit{Agitated smooth transition:}
In the agitated case (matched slopes, $\Delta\alpha=0$, with an offset jump $\Delta(\phi_{\rm c})=\Delta V>0$), one has $\Delta(\phi)\simeq\Delta V$ near the centre. Using $V_{,\phi\phi}=2\,\Delta\alpha\,S'(\phi)+\Delta(\phi)\,S''(\phi)$ from \cref{eq:Vpp_generalS}, the turning–point condition for $H$, $V_{,\phi\phi}=\tfrac{2}{3}$ reduces to
\begin{equation}
\Delta V\,S''(\phi)=\frac{2}{3}.
\label{eq:turning_agitated}
\end{equation}
For the logistic interpolant, $S''(\phi)=-\eta^{2}\sech^{2}u\,\tanh u$ is positive only on the AdS side ($u<0$) and attains its positive maximum at
$u=-\tanh^{-1}\!(1/\sqrt{3})$, where
\begin{equation}
S''_{\max}=\frac{2}{3\sqrt{3}}\,\eta^{2}.
\end{equation}
Thus a necessary (and for the logistic case, also sufficient) condition for interior turning points is $\Delta V\,S''_{\max}>2/3$, i.e.,
\begin{equation}
\Delta V\,\eta^{2}>\sqrt{3}.
\label{eq:threshold_agitated_logistic}
\end{equation}
A canonical matched–slope realisation with flat plateaus is $V(\phi)=-\Lambda_{\rm dS}+2\Lambda_{\rm dS}\,S(\phi)$, for which $V_{,\phi}=\eta\,\Lambda_{\rm dS}\,\sech^{2}u$ and
\begin{equation}
H(\phi)=\frac{\eta}{2}\,\Lambda_{\rm dS}\,\sech^{2}u-\frac{\phi}{3}.
\label{eq:H_agitated_sech2}
\end{equation}
Writing $u=\eta(\phi-\phi_{\rm c})$ and $\phi_{\rm c}=-\zeta/\eta$, one finds
\begin{equation}
H_{,\phi}=-\,\eta^{2}\Lambda_{\rm dS}\,\sech^{2}u\,\tanh u-\tfrac13,
\end{equation}
with operational endpoints
\begin{equation}
\begin{aligned}
H_{\rm on}&=\frac{\eta}{2}\Lambda_{\rm dS}\sech^{2}2+\frac{\zeta+2}{3\eta},\\
H_{\rm ex}&=\frac{\eta}{2}\Lambda_{\rm dS}+\frac{\zeta}{3\eta},\\
H_{\rm end}&=\frac{\eta}{2}\Lambda_{\rm dS}\sech^{2}2+\frac{\zeta-2}{3\eta}.
\end{aligned}
\end{equation}
Interior extrema require $H_{,\phi}=0$, i.e.
\begin{equation}
-\,\eta^{2}\Lambda_{\rm dS}\,\sech^{2}u\,\tanh u=\tfrac{1}{3}.
\end{equation}
With ${y\equiv\tanh u}$ this becomes ${y(y^{2}-1)=1/(3\eta^{2}\Lambda_{\rm dS})}$. Since $\max_{u<0}\!\big[\sech^{2}u(-\tanh u)\big]=2/(3\sqrt{3})$ at $u=-\tanh^{-1}\!(1/\sqrt{3})$, two AdS–side turning points $u_\pm$ exist \emph{iff} $\eta^{2}\Lambda_{\rm dS}>\sqrt{3}/2$, which is equivalent to \cref{eq:threshold_agitated_logistic} with $\Delta V=2\Lambda_{\rm dS}$. In the sharp regime $\eta^{2}\Lambda_{\rm dS}\gg1$,
\begin{equation}
u_{+}\simeq -\frac{1}{3\,\eta^{2}\Lambda_{\rm dS}},\quad
u_{-}\simeq -\frac{1}{2}\ln\!\big(12\,\eta^{2}\Lambda_{\rm dS}\big),
\end{equation}
and inserting into \cref{eq:H_agitated_sech2} yields
\begin{equation}
\begin{aligned}
H_{\max}&\simeq \frac{\eta}{2}\Lambda_{\rm dS}+\frac{\zeta}{3\eta}
+\mathcal{O}\!\big((\eta^{3}\Lambda_{\rm dS})^{-1}\big),\\
H_{\min}&\simeq \frac{\zeta}{3\eta}+\frac{1}{6\eta}\!\left[1+\ln\!\big(12\,\eta^{2}\Lambda_{\rm dS}\big)\right].
\end{aligned}
\end{equation}
For endpoint comparisons, using $\sech^{2}2=1-\tanh^{2}2$ and $u_{+}$ above, one finds
\begin{equation}
\begin{aligned}
H_{\max}-H_{\rm on}&\;\simeq\;\frac{\eta}{2}\Lambda_{\rm dS}\,\tanh^{2}\!2\;-\;\frac{2}{3\eta}
+\mathcal{O}\!\big((\eta^{3}\Lambda_{\rm dS})^{-1}\big),\\
H_{\max}-H_{\rm end}&\;\simeq\;\frac{\eta}{2}\Lambda_{\rm dS}\,\tanh^{2}\!2\;+\;\frac{2}{3\eta}
+\mathcal{O}\!\big((\eta^{3}\Lambda_{\rm dS})^{-1}\big),
\end{aligned}
\end{equation}
so the bump height within the window grows linearly with $\eta\,\Lambda_{\rm dS}$ (geometric prefactor $\tanh^{2}\!2\simeq0.93$), while the overshoot above the centre, $H_{\max}-H_{\rm ex}=\mathcal{O}((\eta^{3}\Lambda_{\rm dS})^{-1})$, is parametrically small. When $\eta^{2}\Lambda_{\rm dS}\le\sqrt{3}/2$, $H_{,\phi}=0$ admits no interior AdS–side solution and $H(\phi)$ remains a gentle, monotone profile across the window, with no genuine interior maximum (no bump).}\\

\noindent The turning–point condition for $H$, viz., $V_{,\phi\phi}=\tfrac{2}{3}$, together with the thresholds \cref{eq:threshold_quiescent_logistic,eq:threshold_agitated_logistic}, collectively provide a compact, model–agnostic diagnostic for the presence and sharpness of a super–acceleration bump. There are two distinct bump drivers:
(i) a \emph{slope–mismatch} contribution $\propto \Delta\alpha\,S'$ (quiescent smooth transition, continuous parent with $\Delta(\phi_{\rm c})=0$),
and (ii) an \emph{offset–jump} curvature contribution $\propto \Delta(\phi)\,S''$ (agitated smooth transition, matched slopes $\Delta\alpha=0$ but $\Delta V>0$).
The former is bounded by the shape of $S'$, whereas the latter can be amplified by the $\eta^{2}$ scaling in $S''$. In the \emph{quiescent} case the driver is the slope mismatch $\Delta\alpha$: the condition $\Delta\alpha\,\eta>2/3$ is necessary, and even deep in the sharp regime the bump amplitude scales essentially linearly with $\Delta\alpha$, with small $1/\eta$ subtractions from the $-\phi/3$ drift, cf.\ \cref{eq:heights_quiescent_asym}. Thus, for mirror AdS–to–dS transitions where $\Delta\alpha$ is fixed by the branch geometry, one does not expect parametrically large bumps. By contrast, in the \emph{agitated} case the curvature spike is controlled by the offset jump $\Delta V$; once $\Delta V\,\eta^{2}>\sqrt{3}$, the bump height within the window grows linearly with $\eta\,\Lambda_{\rm dS}$ (or, more generally, with $\eta\,\Delta V$), cf.\ \cref{eq:H_agitated_sech2}, allowing significantly larger bumps as the transition is sharpened. The explicit formulae above also permit direct endpoint and interior height estimates within the operational transition window $u\in[-2,2]$.

\section{Perturbations analysis}\label{sec:appPert}
At the level of the perturbation equations of motion, it has been shown in~\cite{DeFelice:2020eju} that all the equations, including those for the matter fields, are, in form, identical to those in $\Lambda$CDM. For instance, the shear equation reads
\begin{equation}
    \Psi=\Phi -\frac92\,\frac{a^2}{k^2}\,\sum_I(\varrho_I+p_I)\,\sigma_I\,,
\end{equation}
where we have adopted the \texttt{CLASS}~\cite{2011JCAP...07..034B} notation; for each matter field, $\varrho_I=\rho_I/3\Mpl^2$, $p_I=P_I/3\Mpl^2$, and $k$ denotes the wave number of the modes. This equation is expressed in the Newtonian-gauge invariant fields $\Phi$, $\Psi$, and the shear component of each matter field $\sigma_I$. The only modified equation is:
\begin{equation}
\dot{\Phi}+aH\Psi  = 
\frac{3\,[k^{2}-3a^{2}(\dot{H}/a)]\sum_{I}(\varrho_{I}+p_{I})\,\theta_{I}}{k^{2}\,[2k^{2}/a^{2}+9\sum_{K}(\varrho_{K}+p_{K})]}\,, \label{eqn:dotphi}
\end{equation}
where a dot here represents differentiation w.r.t.\ the conformal time (i.e., $N=a$), and $\theta_{I}$ is the divergence of the $I^{\rm th}$ fluid velocity (see \cite{Ma:1995ey}). The quantity $\dot{H}$ is given in~\cref{eq:dotH} [or, equivalently, in~\cref{eq:dot_H}]. For instance, for a given profile of $\varrho_{\rm s}(a)\equiv \rho_\phi/(3\Mpl^2)$, we have
\begin{align}
    \dot{H}&=\frac{a^2}{2}\,\varrho_{{\rm s},a} -\frac32\,a\sum_I(\varrho_I+p_I)\,,\label{eq:dotH}\\
    \ddot{H}&=-2aH\dot{H}-\frac32\,a\,\dot{p}+2a^3H\varrho_{{\rm s},a}+\frac12\,a^4H\,\varrho_{{\rm s},aa}\,.
\end{align}
Here $\dot{p}=\sum_I\dot{p}_I$, where only the ultra-relativistic species contribute a non-zero value.

In~\cref{eqn:dotphi}, the sums over $K$ and $I$ only include the standard matter fields, excluding the $\phi$-component. As the theory is minimal, there is no extra propagating degree of freedom, and therefore, no additional dynamical equation is required by construction. We observe that ${\dot H}$ deviates from the GR expression only when $V_{,\phi\phi}\neq0$, which also influences the dynamics of the perturbation equations, as evidenced by~\cref{eqn:dotphi}. During a rapid transition epoch in the Hubble parameter, we have shown that at least $|{\dot H}|\gg N H^2$. Consequently, also the perturbations will be non-negligibly affected by this change in dynamics, albeit for a short interval in $z$. This suggests that the mirror AdS-to-dS transition will generally impact the perturbation dynamics as well, and the various types of this transition would leave distinguishing imprints on the perturbations. Since the theory is minimal, the scalar field $\phi$, by construction, is an auxiliary field and therefore does not propagate, preventing it from becoming unstable. In the small-scale regime, all no-ghost conditions for both matter fields and gravitational waves are trivially satisfied. Moreover, the propagation speeds of all modes, including gravitational waves, match those in GR. In the subhorizon regime, the growth of perturbations for the dust field follows the same equation of motion as in GR (see, e.g.,~\cite{DeFelice:2020eju}).

\section{Distinguishing features of the abrupt $\Lambda_{\rm s}$CDM model compared to $\Lambda$CDM}\label{sec:appC}

To clarify how the \textit{abrupt} $\Lambda_{\rm s}$CDM model~\cite{Akarsu:2019hmw,Akarsu:2021fol,Akarsu:2022typ,Akarsu:2023mfb} differs from $\Lambda$CDM, it is critical to note that, unlike the usual $\Lambda$---which remains positive and unchanged throughout cosmic history---$\Lambda_{\rm s}(z)$ becomes negative, $\Lambda_{\rm s}(z) = -\Lambda_{\rm s0}$, for $z > z_{\dagger} \sim 2$. From physical and mathematical perspectives, this alteration extends from the transition epoch at $z_{\dagger} \sim 2$ backward to the early universe, including the recombination era at $z_{\rm rec} \sim 1100$ and beyond. In contrast, for $z < z_{\dagger}$, $\Lambda_{\rm s}$CDM matches $\Lambda$CDM by accommodating a positive CC, $\Lambda_{\rm s}(z) = \Lambda_{\rm s0}$, after the transition---albeit with a larger value than $\Lambda$ in $\Lambda$CDM ($\Lambda_{\rm s0} > \Lambda$) to compensate for its earlier negative phase as explained in~\cite{Akarsu:2021fol}. While one might interpret this negative interval of $\Lambda_{\rm s}(z)$ as an early-time modification from both physical and mathematical standpoints, its main observational impact arises only for $z \lesssim 3$, where $\Lambda_{\rm s}(z)$ switches from negative to positive near $z_{\dagger}$, causing a deformation in $H(z)$ relative to $\Lambda$CDM and thereby yielding a larger expansion rate $H_{\Lambda_{\rm s}\mathrm{CDM}}(z)$ at low redshifts. Consequently, from an observational viewpoint, $\Lambda_{\rm s}$CDM can be regarded as a post-recombination extension of $\Lambda$CDM, classified as a late- or moderate-time modification depending on context. Crucially, for $z \gtrsim 3$, the two models become nearly indistinguishable in both their dynamics and observational signatures, despite their theoretical differences. This similarity is expected because the fractional energy density from $\Lambda_{\rm s}$ (in $\Lambda_{\rm s}$CDM) or $\Lambda$ (in $\Lambda$CDM) comprises only a few percent of the total by $z \sim 3$ and becomes negligible at higher redshifts. In particular, for $3 \lesssim z \lesssim z_{\mathrm{eq}}$ (where $z_{\mathrm{eq}} \sim 3400$ marks matter-radiation equality), both models effectively reduce to an Einstein--de Sitter universe and reproduce the standard cosmological evolution prior to recombination ($z > z_{\mathrm{rec}} \sim 1100$). Notably, the same reasoning holds for perturbations, since $\Lambda_{\rm s}(z)$ only modifies the late-time Hubble rate $H(z)$ and leaves the linear perturbation and Boltzmann equations formally unchanged (assuming GR). Thus, from a physical and mathematical standpoint, one recovers $\Lambda$CDM from abrupt $\Lambda_{\rm s}$CDM by taking $z_{\dagger} \to \infty$, whereas, in terms of the Hubble rate---and hence observationally---$\Lambda_{\rm s}$CDM is effectively indistinguishable from $\Lambda$CDM for $z \gtrsim 3$. This also underscores that choosing $z_{\dagger} \gtrsim 3$ renders $\Lambda_{\rm s}$CDM practically identical to $\Lambda$CDM in observable data.

The dynamics of CDM and baryonic species---including their clustering properties---remain unchanged by design in $\Lambda_{\rm s}$CDM, matching those in (abrupt) $\Lambda$CDM. However, this does not imply that $\Lambda_{\rm s}$CDM leaves the inferred values of cosmological parameters, such as $\Omega_{\rm m0}$ (the present-day matter density parameter) and $H_0$, unchanged in observational analyses. Since the pre-recombination universe is preserved as in $\Lambda$CDM, the comoving sound horizon at last scattering, $r_* = \int_{z_*}^{\infty} c_{\mathrm{s}} H(z)^{-1} \mathrm{d}z$, where $z_* \sim 1090$ is the redshift of the last scattering surface and $c_{\mathrm{s}}$ is the sound speed in the photon-baryon plasma, remains essentially the same as in $\Lambda$CDM. From the Planck CMB spectra, the angular scale of the sound horizon, $\theta_* = r_*/D_M(z_*)$, and the present-day physical matter density, $\Omega_{\rm m0} h^2$ (derived from the peak structure and damping tail, with $h \equiv H_0/100\,\mathrm{km\,s^{-1}Mpc^{-1}}$), are both measured with high precision, largely model-independently. Hence, in $\Lambda_{\rm s}$CDM, both the comoving angular diameter distance to last scattering, $D_{M}(z_*) = c \int_{0}^{z_*} H(z)^{-1} \mathrm{d}z$, and $\Omega_{\rm m0} h^2$ are expected to remain consistent with their Planck-inferred values for $\Lambda$CDM. Nevertheless, any suppression of $H(z)$ at $z > z_\dagger$, due to the negative cosmological constant in this regime, must be offset by an enhancement at lower redshifts ($z < z_\dagger$) to maintain this consistency, thereby increasing $H_0$. Since $\Omega_{\rm m0} h^2$ is tightly constrained by the CMB, a larger $H_0$ necessitates a smaller $\Omega_{\rm m0}$ (as $\Omega_{\Lambda_{\rm s0}} + \Omega_{\rm m0} = 1$, implying a larger $\Omega_{\Lambda_{\rm s0}}$ compared to $\Omega_{\Lambda}$ of $\Lambda$CDM), reflecting the well-known negative correlation in the $H_0-\Omega_{\rm m0}$ plane that also exists in $\Lambda$CDM. Indeed, fits to $\Lambda_{\rm s}$CDM reveal a reduced $\Omega_{\rm m0}$, arising solely from this late-time background deformation in the $H(z)$ function, even though the physics of CDM and baryons remains identical to that in $\Lambda$CDM~\cite{Akarsu:2024eoo}.
\bibliographystyle{apsrev4-2_mod_yearfix}
\bibliography{bib}

\end{document}